# Encapsulation of oils and fragrances by core-in-shell structures from silica particles, polymers and surfactants: The brick-and-mortar concept

Gergana M. Radulova [a], Tatiana G. Slavova [a], Peter A. Kralchevsky [a,*], Elka S. Basheva [a], Krastanka G. Marinova [a], Krassimir D. Danov [a]

[a] *Department of Chemical and Pharmaceutical Engineering, Faculty of Chemistry and Pharmacy, Sofia University, 1164 Sofia, Bulgaria*

ABSTRACT. Colloidosomes provide a possibility to encapsulate oily substances in water in the form of core-in-shell structures. In this study, we produced microcapsules with shell from colloidal particles, where the interparticle openings are blocked by mixed layers from polymer and surfactant that prevent the leakage of cargo molecules. In other words, the particles and polymer play the role of bricks and mortar. For this goal, we used hydrophilic silica particles, which were partially hydrophobized by the adsorption of potassium oleate to enable them to stabilize Pickering emulsions. Various polymers were tested to select the most appropriate one. The procedure of encapsulation is simple and includes single homogenization by ultrasound. The produced capsules are pH responsive. They are stable in aqueous phase of pH in the range 3 – 6, but at pH > 6 they are destabilized and their cargo is released. With the optimized formulation of silica particles, polymer, oleate and NaCl, we were able to encapsulate various oils and fragrances, such as tetradecane, limonene, benzyl salicylate and citronellol. All of them have a limited and not too high solubility in water. In contrast, no stable microcapsules were obtained with oils that either have zero water solubility (mineral and silicone oil) or higher water solubility (phenoxyethanol and benzyl alcohol). By analysis of results from additional interfacial-tension and thin-film experiments, we concluded that a key factor for obtaining stable capsules is the irreversible adsorption of the polymer at the oil/water interface. The hydrophobization of the particles by surfactant adsorption (instead of silanization) plays an important role for the pH responsiveness of the produced capsules. The obtained information about the role of various factors for the stabilization of microcapsules, which are based on the brick-and-mortar concept, can be further used to achieve better stability; selection of polymers that are appropriate for different classes of oils, as well as for the production of smaller capsules stabilized by nanoparticles.

* Corresponding author.
*E-mail address:* pk@lcpe.uni-sofia.bg (P.A. Kralchevsky).

# 1. Introduction.

The development of micro- and nanocapsules with controlled composition and properties has been a subject of intensive research in the last decade. The stimuli responsive release of cargo molecules from the carrier gains a remarkable attention for *in vivo* and *in vitro* delivery of contrast agents, genes, and pharmaceuticals [1]. The colloidal capsules find another broad field of applications in house-hold and personal-care detergency, cosmetics and food industry, e.g. for the encapsulation of fragrances, flavors, colorants and preservatives [2,3].

By their *chemical nature*, the colloidal carriers are classified as inorganic, organic and hybrid, e.g. mineralized protein capsules [1,4]. Different classes of colloidal carriers can be distinguished in relation to their *structure and morphology*: micelles; dendrimers; liposomes; niosomes; polymersomes; cubosomes and hexosomes; colloidosomes, and emulsion-based micro- and nanocapsules [1,5]. The colloidal capsules differ also by the *mechanisms of cargo release* (see below). Because the colloid science offers a rich variety of systems, structures and mechanisms, the potential for discovery and development of novel systems with useful and unique properties is enormous.

Detailed overviews on the achievements in the field of micro- and nanocapsules can be found in recent review articles [1,2,5]. Here, we only briefly present the current state of the art in this area. The colloidal carriers, classified by their structure and morphology, are as follows:

*Micelles* are self-assembled aggregates from amphiphilic molecules, such as surfactants, di-block and tri-block copolymers. With micelles, the control of release of the encapsulated material is not easy [1]. Nevertheless, they have some advantages over the other colloidal carriers, such as small and uniform size; low toxicity; increased cargo solubility and bioavailability, etc. [6,7,8]. *Microemulsion* droplets could execute analogous functions [9,10,11].

*Dendrimers* are polymer-based hyperbranched amphiphilic macromolecules, 1–10 nm in diameter. They can be loaded with cargo molecules owing to the hydrophobic/hydrophilic interactions. Modification of the degree of branching may allow encapsulation of different molecules [12,13].

*Liposomes* represent closed spherical phospholipid bilayers (vesicles) of thickness ca. 3 – 4 nm. They are able to solubilize lipophilic substances in the phospholipid bilayer. Moreover, a hydrophilic cargo can be loaded in the water cavity of the liposomes. Thus, they can serve as carriers of two cargos and can protect them from environmental influence. Liposomes have been used for encapsulation of therapeutic agents [14], for drug delivery [15], and in combination with the layer-by-layer (LbL) technique [16].

*Niosomes* have a structure, which is similar to that of liposomes, but instead of phospholipid molecules, the structural units are surfactant molecules. They have similar functions like the liposomes [17,18]. The niosomes are usually built up of single-chained nonionic surfactants, like Span 60, but they can be also a result of self-assembly of gemini surfactants [19].

*Polymersomes* also represent bilayer vesicles, but composed of block copolymers. In general, the polymersomes are more stable as compared to liposomes and niosomes. Polymersomes have been used to deliver encapsulated anti-cancer drugs [20,21]. Magnetic polymersomes are of interest as nanocarriers for monitoring of drug delivery [22].

*Cubosomes* and *hexosomes* are nonlamellar liquid crystalline phases, which offer attractive potential platforms for cargo solubilization and targeted delivery [5,23,24,25]. The cubosomes are bicontinuous cubic phases, which could solubilize both hydrophobic and hydrophilic cargo molecules. The hexosomes are reverse hexagonal phases. Biologically active molecules can either be accommodated within the aqueous domains of the hexosomes or can be directly coupled to the lipid hydrophobic moieties oriented radially outwards from the center of the water rods [5].

*Colloidosomes* are shell-like supra-particles composed of small colloidal spheres first produced by using emulsion drops as templates [26]. The term colloidosomes was coined later [27]. Colloidosomes have been used to encapsulate various substances [28-31] and even biological cells [32,33]. The electrostatically stabilized colloidosomes composed of positively and negatively charged colloid particles represent interesting class of carriers [34,35]. The colloidosomes can be disassembled and the cargo released by thermal control [36]; pH control [37,38,39], and even by light-triggered disassembly [34].

*Emulsion-based micro- and nanocapsules* are prepared by emulsification in turbulent flow, by membrane emulsification or by various micro-fluidic techniques. The cargo is located in the disperse phase, i.e. in the droplets, which are stabilized against coalescence by appropriate amphiphilic molecules, which can be surfactants, polymers and proteins [40,41]. Warszynski et al. developed a methodology for producing microcapsules [42] and nanocapsules [43,44,45], which is based on emulsion drops that are covered by multilayered polyelectrolyte shells (composed of layers from anionic and cationic polymers) obtained by the LbL technique.

With respect to the mechanisms of cargo release, one could distinguish *passive* release and *controlled* release. The passive release happens by diffusion through the capsule and/or by its degradation (bioerosion). In the case of controlled release, several *stimuli-responsive* release mechanisms exist in relation to the character of the external triggers: (i) changes in pH; (ii) changes in temperature; (iii) changes in redox potential; (iv) action of enzymes; (v) action of magnetic field, and (vi) light irradiation; for details, see Ref. [1].

Despite the great advance in the field of design and development of colloidal carriers for biotechnology, nanomedicine, cosmetics and food industry, there is a vast field for further developments in view of the great choice of materials and stabilizing agents for formulation; methods for preparation, and ways for release control.

As already mentioned, the encapsulation of an oily phase by colloidosomes can be achieved by production of particle-stabilized Pickering emulsion [26]. A conventional colloidosome releases its cargo (e.g. fragrance) by a slow dissolution through the openings between the particles in the colloidal shell, i.e. by passive release. It is not easy to achieve controlled release by destruction of the Pickering emulsion, because such emulsions are rather stable [46,47]. The production of composite microcapsules formed from solid particles and polymers typically contain an emulsion-polymerization step including crosslinking by chemical bonds [48,49].

Our aim in the present study is to demonstrate that stable composite capsules from particles and polymers can be formed under the action of weaker physical (colloidal) forces, without any polymerization step. pH-responsive colloidosome-based composite microcapsules are developed, which are loaded by oils, including fragrances. In the case of conventional colloidosomes, there can be passive cargo release through the interparticle openings. To prevent such uncontrollable leakage, we blocked the openings between the particles in the colloidal shell by a mixed adsorption layer of polymer and surfactant, which resembles a brick-and-mortar structure. The role of surfactant is to provide reversible hydrophobization of the used silica particles. The role of the particles (the bricks) is related to their irreversible attachment to the surface of the oil drops, thus, creating a colloidal shell – a basic element of the capsule. Inorganic electrolyte was also used to suppress electrostatic barriers that hinder the particle entry at the oil/water interface. The factors that have been varied in our experiments to improve the capsule stability are the type and concentration of electrolyte; surfactant, and polymer. After selecting the optimal composition, the type of oil was varied as well. In the cargo-release experiments, the pH and temperature were varied. Upon self-storage, the obtained microcapsules (in aqueous medium) are stable for at least 8 months – the period of our observations. However, upon raising the pH above 6, these capsules are quickly destroyed and release their cargo.

In Section 2, the used materials and methods are described. In Section 3, we report our results on encapsulation of various oils by optimizing the composition of the used aqueous suspension. In Section 4, we present the results from additional experiments with thin emulsion films, indicating the experimental conditions for which the polymer adsorbs on the oil-water interface; whether the surfactant could displace the polymer from the interface, and whether there are pronounced surfactant-polymer interactions. Based on these results, in Section 5 we discuss the roles of the main factors that govern the formation of stable or unstable microcapsules.

**2. Materials and methods**

*2.1. Materials*

The used $SiO_2$ particles Excelica UF320 were product of Tokuyama Corp. (Japan). They are high-purity synthetic spherical fused silica beads produced from silicon tetrachloride in a gas phase reaction. The particles are hydrophilic of density of 2.2 g/cm$^3$ and have been used without any further modification. They are perfectly spherical, with smooth surface and average diameter $d = 3.5$ μm. Their size distribution is lognormal with mean dispersion $\sigma = 1.23$ μm. Consequently, 50% of the particles have diameter $d_p$ in the interval $d/\sigma \leq d_p \leq \sigma d$ that is $2.8 \leq d_p \leq 4.3$ μm; see Fig. A1 in the Appendix and Ref. [50].

The organic liquids that were subject to encapsulation were limonene (product of Sigma Aldrich); benzyl salicylate (Sigma Aldrich); lilial (ABCR GmbH&Co. KG); citronellol (Sigma Aldrich); tetradecane (Sigma Aldrich); sunflower seed oil (SFO, from a local supplier); oleic acid (TCI); light mineral oil (LMO, Sigma Aldrich); silicone oil AK 10 of kinematic viscosity 10 cSt (Wacker); benzyl alcohol (Fisher); phenoxyethanol and linalool (products of Sigma Aldrich).

The LMO and tetradecane are representatives of the mineral oils; SFO and AK 10 are representatives of the vegetable and silicone oils, respectively. The oleic acid represents the liquid fatty acids. Limonene, lilial, linalool, citronellol and benzyl salicylate are fragrances used in personal care (cosmetic) products. Benzyl alcohol and phenoxyethanol are also used in personal care products, but as bacteriostatic preservatives.

The typical sunflower seed oil (SFO) is composed of triglycerides derived from the following main fatty acids: 52 wt% linoleic acid (polyunsaturated omega-6); 34% oleic acid (monounsaturated omega-9), and 12 wt% palmitic acid (saturated) [51]. The SFO was purified from other admixtures by passing through a column filled with Silica Gel and Florisil adsorbent. The measured value of the interfacial tension against pure water was 31 mN/m, a typical value for pure SFO [52], and it did not decrease by more than 0.2 mN/m within 60 min. The other oils were used without any further purification. For better visualization, we dissolved 0.12 wt% of the dye Sudan III (Sigma Aldrich) in the oily phase. Thus, the oil drops acquire a reddish color and were better visible.

The used inorganic salts were sodium chloride (NaCl, product of Sigma Aldrich), potassium chloride (KCl, Merck), zinc chloride ($ZnCl_2$, Fluka), calcium chloride ($CaCl_2$, Chemlab), iron sulfate ($FeSO_4$, Sigma Aldrich), and aluminum chloride ($AlCl_3$, Sigma Aldrich). As stabilizing surfactants, we used carboxylates: sodium laurate (NaLaurate, TCI); potassium myristate (KMyristate, Viva Corporation, India), and potassium oleate (KOleate, Fluka), of hydrocarbon chainlengths C12, C14 and C18, respectively. The longer chainlengths are expected to provide stronger hydrophobization of the silica particles.

The following polymers were used: Carbopol® 971P NF; Carbopol® 974P NF; Carbopol® 980 NF, and Carbopol® ETD 2020 NF, all of them products of Lubrizol. The first three Carbopols are polymers of acrylic acid crosslinked with allyl sucrose or allyl pentaerythritol, whereas Carbopol ETD 2020 is a block copolymer of polyethylene glycol and a long chain alkyl acid ester. Other used polymers were polyacrylic acid (PAA, Acros organics); Natrosol$^{TM}$ 250 HHR (Ashland, a hydroxyethylcellulose polymer); carboxymethylcellulose (CMC, Sigma Aldrich); Darvan® 670; Darvan® 7-NS and Darvan® 811-D, all of them products of Vanderbilt Minerals LLC, and Jaguar® S (Solvay). Darvan 670 is sodium polynaphthalene sulfonate; Darvan 7-NS is sodium polymethacrylate and Darvan 811D is sodium polyacrylate. Jaguar S, that is Cyamopsis Tetragonoloba (Guar) Gum, is a native polymer of very high molecular weight without any chemical modifications.

Insofar as the best results have been obtained with the Carbopols, in Table A1 of the Appendix we give the viscosity ranges of Carbopol solutions (as provided by the producer) and the zeta potentials of Carbopol aggregates in water (measured by us). From the data in Table A1, it follows that Carbopol 971P NF is the polymer of the lowest molecular mass and the lowest degree of crosslinking among the used Carbopols.

In some experiments we used also the surfactants polyoxyethylenesorbitan monolaurate (Tween 20, Sigma Aldrich), sodium laurylethersulfate with one ethylene oxide group (SLES-1EO) product of Stepan Co., and sulfonated methyl ester with lauryl chain (C12-SME) product of KLK-Oleomas (Malaysia). At sufficiently high concentrations, these detergents (of alkyl chainlength C12) are expected to be able to destroy the capsules and to release the cargo.

All aqueous solutions were prepared with deionized water (Elix purification system, Milipore). In some test experiments, the pH was decreased to 3 – 3.5 by adding HCl, or increased to 10 – 10.5 by adding NaOH. In other experiments, we added blue dye (BD, Brilliant Blue FCF) or methylene blue (MB), both of them products of Sigma Aldrich, to the aqueous phase to distinguish it better from the reddish oily phase. All experiments were carried out at a room temperature of 25 ºC.

*2.2. Methods*

*Preparation of dispersions.* The procedure for production of stable microcapsules, which was established by trials and errors, is the following (Fig. 1). First, we placed the silica particles in the working vial. Next, aqueous solution of carboxylate (KOleate in most experiments) was added and the vial was gently shaken by hand to disperse the particles. The carboxylate concentration (1 – 3 mM) was high enough to ensure a moderate particle hydrophobization. (At higher surfactant concentrations, the hydrophobized particles began to form aggregates in water and even to enter the oily phase; see Fig. A3 in the Appendix.) Next, water solution of polymer and inorganic electrolyte (e.g. NaCl) was added and the dispersion was stirred by hand shaking. Last, we added the oily phase. The total volume of the working samples was 10 mL. They contained 5 wt% (0.5 g) silica particles, 10 vol% oil (e.g. 0.84 g limonene) and the rest (e.g. 8.66 g) was water solution.

Finally, to prepare the dispersion the vial was placed in an Ultrasonic Pulse Homogenizer SKL1500 – IIDN. The sonication was conducted with a sonotrode of diameter 3.24 mm by 1 s long pulses of power output around 200 W, followed by 0.5 s off time.

*Rinsing the microcapsules with water.* After the sonication, the vial with the dispersion was left for 30 min at rest. During this time, the microcapsules with the oily phase, which were covered with silica particles and were heavier than water, sedimented on the bottom of the vial (Fig. 1). After that, the water phase above the sediment was carefully removed with a syringe. This water phase contained a mixture of particles, polymer and carboxylate, which were not adsorbed on the surface of the oil drops. Next, pure water was poured over the microcapsules and the bottle was gently shaken by hand. The sample was left for 30 min at rest. After that, the rinsing was repeated once again. The goal was to remove the excess amounts of carboxylate, polymer, NaCl, and non-adsorbed particles thus making the microscopic observations easier. We call *stable microcapsules* those, which survive after the described procedure of rinsing.

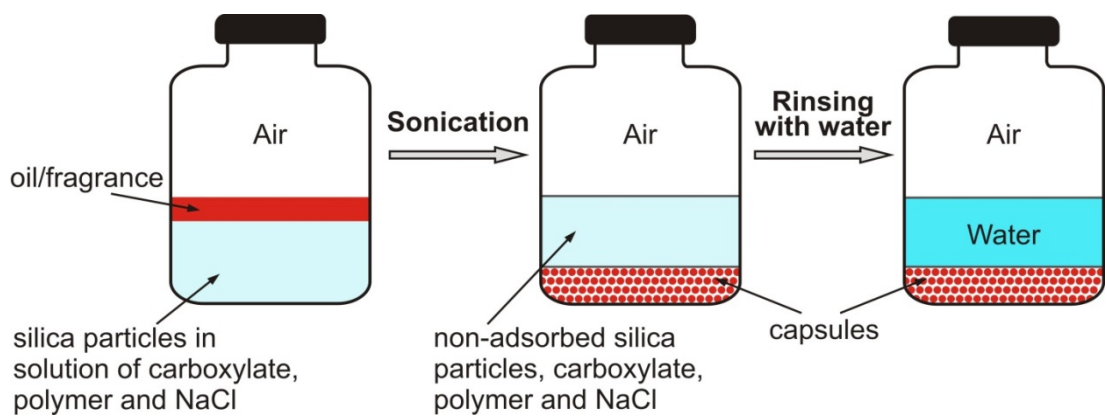

**Fig. 1.** Scheme of the processes of microcapsule preparation and rinsing; details in the text.

*Microscopic observations.* To observe the result of sonication and rinsing, the obtained dispersions were studied by microscope. A small volume of the sediment (Fig. 1) was taken by a pipette and a drop was placed on a microscope slide. On the top of this drop, a cover slip was placed and the sample was observed in transmitted light. We used Axioplan microscope (Zeiss, Germany), equipped with Epiplan objectives (10×, 20× and 50×) and connected to CCD camera with digital recorder. When closing the microcapsule suspension between the microscopic slide and the cover slip, the rigid capsules do not deform – they preserve their spherical shape. With some formulations, we observe less rigid microcapsules, which deform when the concentrated suspension is pressed between the slide and the cover slip (see below).

*Interfacial tension measurements.* The oil/water interfacial tension (IFT) was measured by a DSA 10 apparatus (Krüss GmbH, Germany) using the pendant/buoyant drop method. For various oils, the IFT was measured against both pure water and the working solution, which could contain surfactant, polymer and salt.

*Adjustment of pH.* To verify whether the produced microcapsules are pH-sensitive, we carried out experiments also at acidic and basic pH values. For this goal, the rinsed samples were loaded in pH-STAT Titrando 842 (Metrohm, Switzerland) and their pH was adjusted by the addition of HCl or NaOH. During the experiment, the sample was subjected to gentle stirring with a magnetic stirrer.

*Experiments with thin liquid films.* In the experiments with individual free emulsion films, the Scheludko–Exerowa (SE) cell [53] (called also the capillary cell) was used. It consists of a glass capillary of inner radius 1.5 mm with a side capillary, which is used to inject/suck-out the film phase. A detailed description of the application of SE cell for study of emulsion films can be found in Ref. [54]. In the case of aqueous films, the inner wall of the capillary is hydrophilic. Foam films are formed by filling the capillary by the investigated aqueous solution through the side capillary, with subsequent sucking-out of the liquid, which leads to the formation of a plane parallel film in the central part of the cell. Emulsion films are formed in a similar way with the main difference being that the cell is initially filled with the oil (instead of air) phase [54,55]. The film is observed in reflected light with illumination through the objective of the microscope. Very thin films (of thickness < 30 nm) look black because of the very low intensity of the reflected light. At greater film thicknesses, the films look brighter. In monochromatic light, above the first interference maximum (at thickness > 100 nm), a series of interference minima and maxima are observed. In polychromatic light, one observes interference fringes of various colors. In our experiments, the SE cell was used to detect the presence of polymer aggregates (~100 nm in diameter) adherent to the surfactant adsorption layers at the film surfaces and arrested in the film. In the case of unstable films, the film lifetime (between the film formation and rupture) was also measured.

## 3. Experimental results on encapsulation

Here, we first report results for the effects of inorganic electrolyte, surfactant and polymer on the encapsulation. As oil phase, limonene was used in this series of experiments. The choice of limonene was because of its applications as a dietary supplement and as a fragrance ingredient for cosmetics products. As the main fragrance of citrus peels, limonene is used in food manufacturing, in some medicines, and as a fragrance in perfumery, aftershave lotions, bath products, and other personal care products.

After establishing the optimal composition of the aqueous phase for obtaining stable microcapsules, experiments with different oil phases were carried out. Furthermore, we investigated whether the produced microcapsules, which are stable at $T$ = 25 ºC and pH = 5.5 – 6 (of the water phase used to rinse the microcapsules), are sensitive to the variation of temperature and pH.

*3.1. Effect of inorganic electrolyte*

When the sample consisted of silica particles dispersed in pure water and limonene, the sonication led to the formation of limonene drops that were covered with particles. However, the produced emulsion was very unstable – even the lightest hand-shaking of the vial with the dispersion led to a visible destruction of the emulsion and release of the encapsulated oil. One possible reason for this instability could be the low surface density of particles attached to the drop surface, which are insufficient to provide steric (Pickering) stabilization.

To suppress possible electrostatic barriers to particle attachment to the oil/water interface [56], we added inorganic electrolyte in the water phase. The concentration needed for the formation of microcapsules depends on the valence of the metal ion – higher concentrations are needed for the ions of lower valence. Thus, in the case of monovalent cations, we used 400 mM NaCl or KCl (Fig. 2A), whereas for $Ca^{2+}$ we used 3.7 mM $CaCl_2$ (Fig. 2B). In the case of $Al^{3+}$, 1 mM $AlCl_3$ was enough to obtain microcapsules covered with particles; see Fig. A2 in the Appendix. As seen in these figures, if inorganic salt was present in the sample, microcapsules were formed. However, their stability was not improved very much – again, shaking of the sample by hand led to microcapsule destruction.

It should be noted that the micrographs shown in Fig. 2 (and other similar figures in the paper) have been taken in transmitted light. For this reason, the color (due to the dye Sudan III in the oil) is more intensive for the bigger drops; less intensive for the smaller drop, whereas the small transparent beads are the silica particles. In some micrographs, small aggregates of particles are seen in the neighborhood of the capsules.

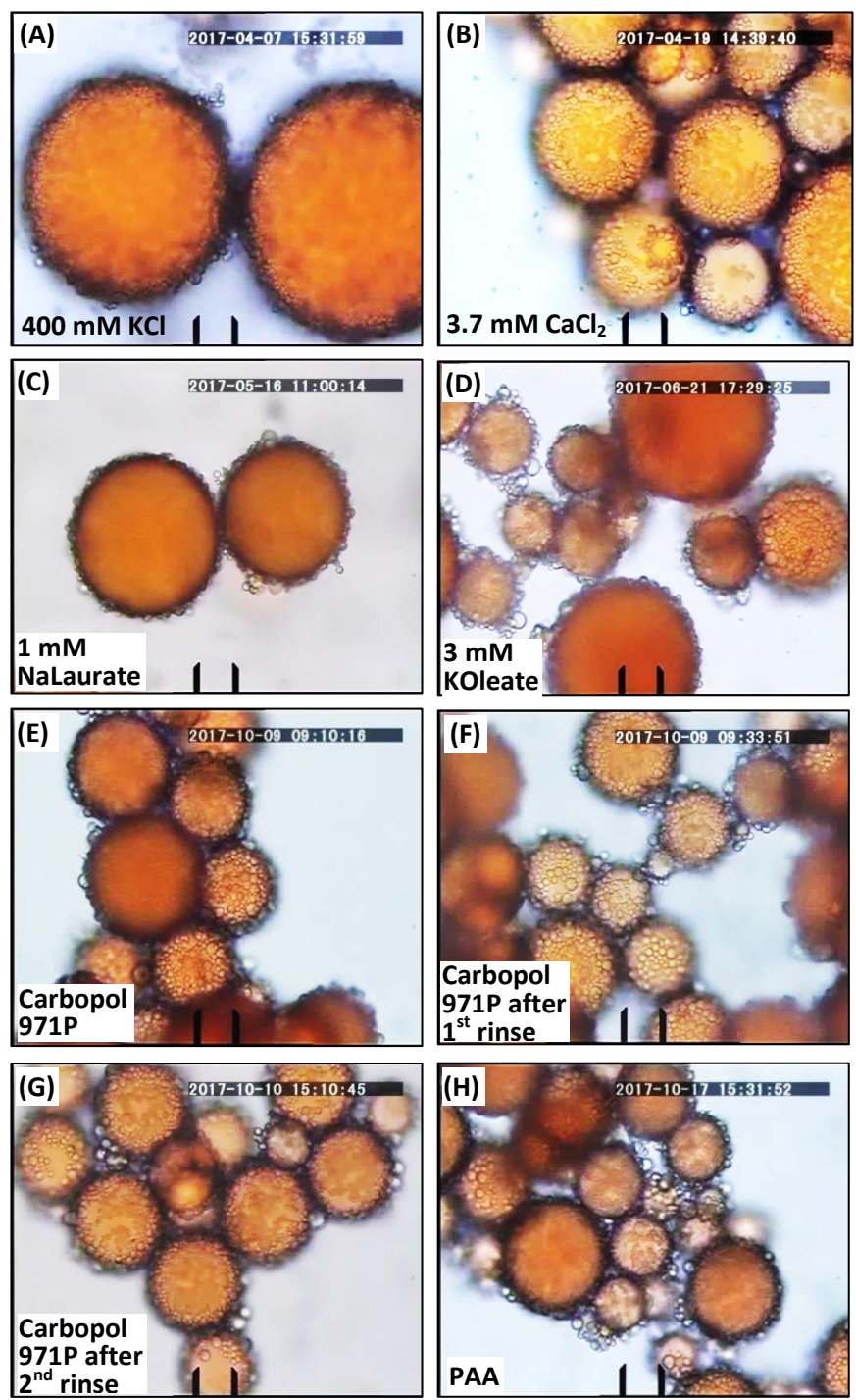

**Fig. 2.** Micrographs of the produced microcapsules: (A) with 400 mM KCl; (B) with 3.7 mM CaCl$_2$; (C) with 1 mM Na Laurate; (D) with 3 mM K Oleate; (E) with Carbopol 971P; (F) with Carbopol 971P after 1$^{st}$ rinse with water; (G) with Carbopol 971P after 2$^{nd}$ rinse with water, and (H) with polyacrylic acid (PAA). The scaling mark is 20 μm.

*3.2. Effect of added carboxylate*

We tried to improve the stability of the microcapsules by adding carboxylate to the water phase, which contained 500 mM dissolved NaCl. Experiments with sodium laurate, potassium myristate and potassium oleate were carried out.

The carboxylate is expected to adsorb with its headgroup to the particle surface and to render the particles more hydrophobic. The increase of the particle/water/oil contact angle leads to a stronger particle attachment to the oil/water interface and is expected to produce a stabilizing effect on the formed microcapsules. The particles should not be too hydrophobic – otherwise they may aggregate in the water phase or even to enter the oil phase. In such cases, stable microcapsules of oil in water cannot be formed.

Fig. 2C and D show stable microcapsules of limonene formed in aqueous solutions of 1 mM NaLaurate and 3 mM KOleate, respectively. Fig. A3 in the Appendix shows that the addition of 5 or 10 mM NaLaurate leads to the formation of particle aggregates rather than microcapsules, whereas in the presence of 5 mM KMyristate or KOleate the particles have become so hydrophobic that they have entered the oil phase; see the rightmost micrographs in Figs. A3b and A3c in the Appendix. We recall that the particles had been initially dispersed in the aqueous phase, together with the surfactant (carboxylate), but after the oil phase was added and the dispersion was stirred by sonication, we observed that the particles had been transferred to the oil phase.

The best results from our experiments with different carboxylates at various concentrations were obtained in the presence of 3 mM KOleate in the aqueous phase. In this case, the drop coverage with particles was the highest. The microcapsules obtained in the presence of KOleate were the most stable in time in comparison with the other two carboxylates. However, upon rinsing of these microcapsules with water (as described in Section 2.2), the particles were detached from the drop surfaces and the emulsion was destroyed.

It should be noted that the contact angle (measured across water) of the original hydrophilic $SiO_2$ particles on the oil/water interface was $\theta \approx 0$. When we filled a vial in half with water and in half with oil, and putted particles in the oil phase, they sank under the action of gravity; cross the oil/water interface and entered the water phase. In contrast, as already mentioned, in the presence of 5 mM KMyristate or KOleate the particles were observed to enter the oil phase, which indicates a large contact angle, $\theta \to \pi$ (180°). Hence, upon the rise of carboxylate concentration the particle/water/oil contact angle increases from $\theta \approx 0$ to $\theta \approx \pi$.

The limiting values of the normal external force, which a particle attached to the oil/water interface can withstand, are

$$F_\text{w} = 2\pi\gamma R \sin^2\frac{\theta}{2} \quad \text{and} \quad F_\text{oil} = 2\pi\gamma R \cos^2\frac{\theta}{2} \tag{1}$$

where $\gamma$ is the interfacial tension; $R$ is the particle radius; $F_\text{w}$ and $F_\text{oil}$ are the magnitudes of the pulling forces toward water and oil phases, respectively; see e.g. Ref. [57], Eq. (4.6) therein. For normal external force of magnitude $F_\text{N} > F_\text{w}$ (or $F_\text{N} > F_\text{oil}$) the particle detaches and enters the water (oil) phase. Upon rinsing of the capsule suspension, $\theta \to 0$ and the external force for detachment tends to zero as $F_\text{w} \propto (\theta/2)^2$; see Eq. (1). This can explain the capsule destruction with particle detachment from the drops under the action of viscous forces due to hydrodynamic fluxes during the rinsing with water.

At sufficiently high carboxylate concentrations (e.g. 5 mM KMyristate or KOleate), $\theta \to \pi$ and the external force for detachment tends to zero as $F_\text{oil} \propto [(\pi - \theta)/2]^2$; see Eq. (1). This can explain the entry of the hydrophobized particles in the oil phase under the action of the strong viscous forces due to the sonication upon dispersion preparation (Figs. A3b and A3c in the Appendix). Because strong hydrodynamic forces are supposed to act during the sonication, the particles could detach even if $\theta$ is not so close to $\pi$.

*3.3. Effect of added polymer*

In this series of experiments, the aqueous phase contained 3 mM KOleate; 500 mM NaCl, and 0.04 wt % polymer. As before, the oil phase was limonene. The microcapsules were prepared using the procedure described in Section 2.2. All of them were rinsed twice with water of pH = 5.5 – 6 to remove the excess KOleate, polymer and NaCl. The obtained results can be summarized as follows.

With Carbopol 971P, 974P and 980, the formed microcapsules were relatively monodisperse, of mean diameter 21.3 μm with standard deviation 6.1 μm obtained from the measured diameters of 500 capsules. As seen in Fig. 2E, the surfaces of the oil drops are covered with a densely packed layer of silica particles. These microcapsules were very stable upon rinsing with water (Fig. 2F and G) or upon shaking by hand the vial with the suspension.

These are the most stable microcapsules observed in our study. Without polymer in the aqueous phase, such stable microcapsules cannot be obtained.

With PAA, stable microcapsules were formed, but their size distribution was more polydisperse as compared with those obtained with Carbopol 971P, 974P and 980 (Fig. 2H). The microcapsules with PAA were also stable upon rinsing with water with pH = 5.5–6.

With Jaguar S, stable microcapsules that can survive rinsing were obtained. However, with this polymer many microcapsules of irregular, elongated shape were observed. This indicates that the adsorption layer possesses surface shear elasticity, which can be explained with the formation of an elastic network from adsorbed polymer molecules. For Jaguar S and for the other used polymers, illustrative micrographs of the obtained dispersions are shown in Fig. A4 of the Appendix.

In contrast, with Carbopol ETD 2020 the silica particles form separate aggregates, whereas the emulsion drops were not covered by particles. No microcapsules were formed in the presence of Carbopol ETD 2020.

With Darvan 7-NS and Darvan 811-D, we observed the formation of microcapsules, but their stability was not good. During the microscope investigation of the microcapsule suspension, 5-6 minutes after the sample loading we observed destruction of some microcapsules and release of a part of the encapsulated oil.

With Darvan 670, the obtained microcapsules were mechanically unstable – even the slightest shaking of the vial caused their destruction. A small number of microcapsules survived after shaking – their surfaces were densely covered with silica particles. These microcapsules-survivors were seen on a background of a layer of free silica particles released by the destroyed microcapsules; see Fig. A4 in Appendix A. With Natrosol 250 HHR, the microcapsules were also unstable – they broke down within several minutes. With CMC, the stability of the microcapsules was similar to that with Natrosol 250 HHR. During the microscopic observations, we saw that many of the microcapsules were flattened (and some of them – destroyed) within several seconds. After 5 min, very few intact microcapsules had survived.

In summary, the most stable spherical microcapsules which survive rinsing and do not break down when sandwiched between the microscope slide and the cover slip, are formed with PAA and Carbopols 971P, 974P, and 980. Jaguar S also forms stable microcapsules but their shape is non-spherical – elongated or irregular. Among the investigated Carbopols, only Carbopol ETD 2020 does not form microcapsules; this polymer was not used in our subsequent experiments. With the rest of the studied polymers, we did not observe any pronounced improving of the capsule stability in comparison with the situation without polymers. Most of the subsequent experiments were carried out with Carbopol 971P – the polymer that provides the best stability of the microcapsules.

To check the shelf stability of the produced microcapsules, one of the rinsed samples with encapsulated limonene, stabilized by Carbopol 971P, was kept in pure water of pH = 5.5 for 8 months at room temperature. After that, we found that the microcapsules are completely stable, without any visible changes; see Figs. A8 and A9 in the Appendix. In the framework of the experimental accuracy, the average diameter of the capsules was the same, $21 \pm 6$ μm, in the initial suspensions and after 8 months of storage. Analogous results were obtained with PAA, instead of Carbopol 971P.

We carried out also test experiments for preparation of microcapsules using the same protocol and formulation (with Carbopol 971P), but without the surfactant (KOleate). In this case, capsules were not produced at all. The particles aggregated in big lumps of irregular shape, which contained incorporated small oil drops, whereas the predominant amount of oil was separated as a layer above the water phase. This fact indicates the presence of particle-polymer interaction. In the absence of surfactant the polymer cannot stabilize the emulsion drops. However, their production and stabilization against coalescence is the first step toward microcapsule formation.

*3.4. Effect of rinsing with surfactant solutions*

The obtained stable microcapsules with PAA and Carbopols 971P, 974P, and 980 were subjected to physicochemical disturbances to test their stability. As before, all samples contained 5 wt% silica particles; 10 vol% limonene; 500 mM NaCl; 3 mM KOleate, and 0.04 wt% polymer. As in every encapsulation study, we are interested not only in the preparation of stable microcapsules, but also in their destruction and cargo release.

As already mentioned, the rinsing with water of pH = 5.5 – 6 does not destroy the microcapsules formed in the presence of these Carbopols and PAA. When the microcapsules were subjected to a second rinse, we observed a few spots free of particles on the surface of the emulsion drops. Nevertheless, the microcapsules remained stable and the encapsulated limonene was not released in the water phase and separated as a layer at the top of water.

However, if the microcapsules were rinsed with water that contains surfactant of sufficiently high concentration, the microcapsules could be destroyed. Details are presented below.

*Tween 20 and C12-SME.* When rinsed with 1 mM solution of the nonionic surfactant Tween 20 or of the anionic surfactant C12-SME, the microcapsules remained intact and covered with particles. However, rinsing at a higher surfactant concentration, viz. 10 mM Tween 20 or C12-SME, led to microcapsule destruction and release of limonene. At the latter surfactant concentration, most of the particles were detached from the surfaces of the oil drops, which were observed to coalesce with one another.

*SLES-1EO.* When rinsed with 0.01 mM solution of the anionic surfactant SLES-1EO, the microcapsules remained intact and covered with particles. However, rinsing of the microcapsules with 0.1 mM solution of SLES-1EO led to detachment of most of the particles from the surfaces of the oil drops. In this case, limonene was not separated – a common surfactant-stabilized emulsion was formed (instead of a particle stabilized Pickering emulsion). The results show that SLES-1EO displaces the particles from the drop surfaces and destroys the microcapsules at a hundred times lower concentration than Tween 20 and C12-SME.

*3.5. Effect of pH*

The composition of the initial dispersion was the same as in Section 3.4. As usual, after the microcapsules were produced, they sedimented at the bottom of the vial and the supernatant was removed. Next, the particles were rinsed with water of different pH: 3 – 3.5, 5.5 – 6 and 10 – 10.5. The results are illustrated in Fig. 3.

At the acidic pH, 3 – 3.5, the microcapsules were stable and the drops were covered with a closely packed particle layer. Only with PAA we observed the formation of bigger and less rigid microcapsules, as compared to the case with Carbopols. At pH = 5.5 – 6, the microcapsules were also stable, as demonstrated above.

The alkaline pH, 10 – 10.5, destabilizes the microcapsules (Fig. 3). The results depend on the kind of the used polymer. With Carbopol 971P, most of the microcapsules remained stable, although we observed the appearance of spots free of particles on their surfaces and separation of oil lenses on the surface of water. With Carbopol 974P, Carbopol 980 and PAA, the alkaline pH destroyed the microcapsules – we observed separation of limonene as a floating layer on the surface of water, whereas the heavier silica particles sedimented at the bottom of the vial. Emulsion drops without adsorbed particles could be seen in the bulk. We could hypothesize that the different behavior of Carbopol 971P is related to its lower molecular mass and lower degree of crosslinking as compared to Carbopol 974P and Carbopol 980; see Table A1 in the Appendix. The results indicate that these properties of Carbopol 971P diminish its desorption at pH ≈ 10, at which the electrostatic repulsion is enhanced.

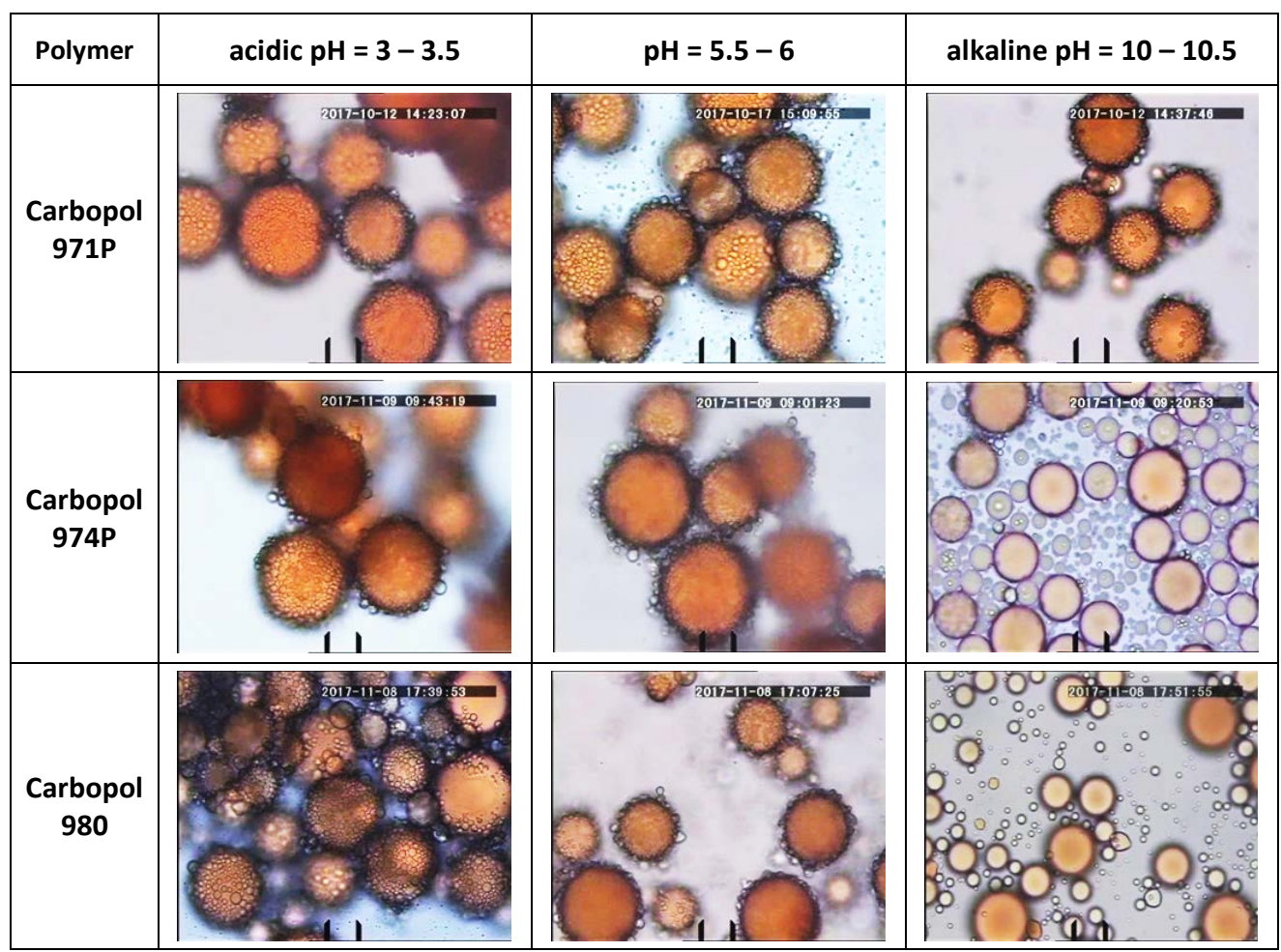

**Fig. 3.** Effect of pH on the properties of microcapsules stabilized by three different polymers after rinsing with water of at three different values of pH, as denoted in the figure. The scaling mark is 20 μm.

To detect more accurately the pH value, at which the capsule destabilization begins, we carried out titration experiments. The experiment started at pH = 2, which was adjusted by the addition of HCl to the suspension with microcapsules. Next, the pH was increased by small steps by the addition of NaOH. At that, the sample was subjected to gentle stirring with magnetic stirrer. For all studied polymers (Carbopol 971P, 974P, 980 and PAA), we observed that the microcapsules began to release limonene at pH = 6 – 7.

Thus, it turns out that the produced microcapsules are *pH-responsive*: they are stable for pH < 6, but they can be destroyed by increasing the pH above 6.

The micrographs of the drops at pH ≈ 10 seen in Fig. 3 are taken soon after the rise of pH by the added NaOH solution. These drops do not immediately coalesce, because they are protected by the negative charge of the oil/water interface at high pH [58]. However, at pH ≈ 10 the emulsions did not possess long-time stability – the thickness of the layer of separated oil phase on the top of solution was increasing with time.

*3.6. Effect of temperature*

The composition of the initial dispersion was the same as in Section 3.4. As polymer, Carbopol 971P was used. The prepared dispersions were rinsed with water of pH in the three ranges denoted in Fig 3. After that, the glass vials with the samples were put for one hour in water bath of temperature 80 ºC. Finally, the dispersions were cooled down to room temperature and studied by microscope. The results, which are illustrated in Fig. A5 in the Appendix, indicate that the heating at 80 ºC does not essentially affect the stability of the studied microcapsules.

In a complementary series of experiments, the dispersions with the microcapsules were frozen (kept in a freezer overnight). Next, the samples were kept at room temperature until reaching 25 ºC. For all the three pH ranges, complete destruction of the microcapsules was observed. The freezing of water at low temperatures destroys the microcapsules.

*3.7. Effect of the kind of oil*

In in these experiments, the composition of the dispersion was the same as in Section 3.4, with the only difference that other oils were used (instead of limonene). The polymer was Carbopol 971P. With this composition of the aqueous phase, microcapsules were formed from sunflower oil; tetradecane; limonene; benzyl salicylate; lilial, and citronellol. Illustrative micrographs are shown in Fig. 4. Details are following.

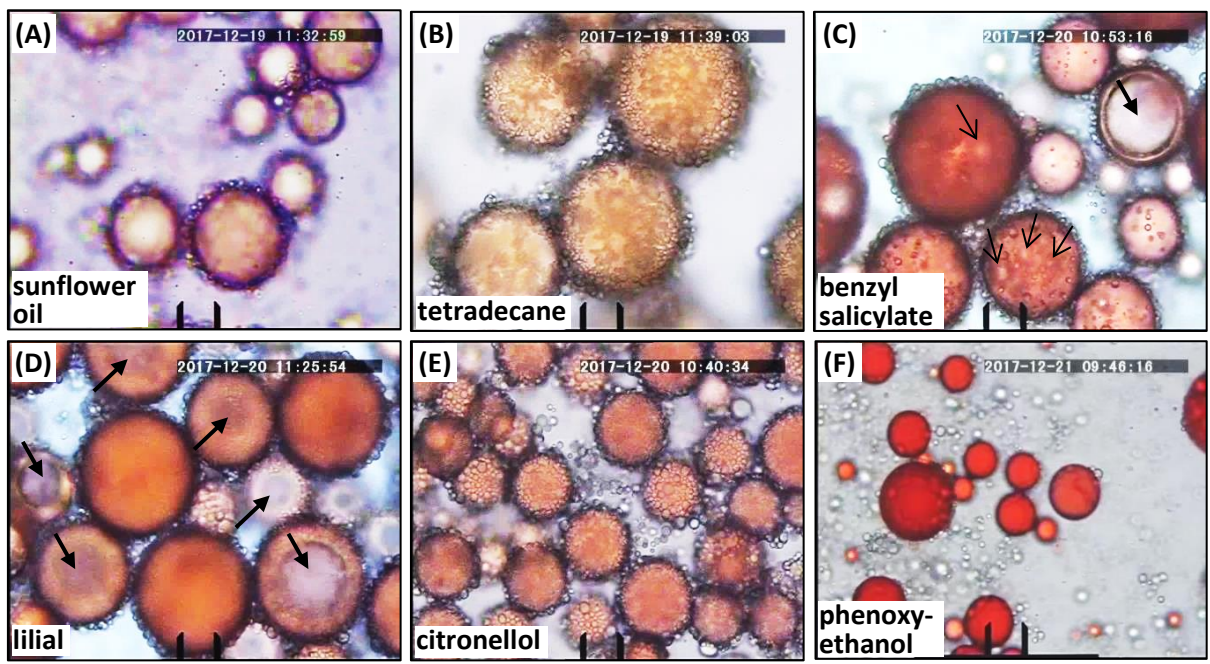

**Fig. 4.** Micrographs of microcapsules of different oils/fragrances stabilized by Carbopol 971P: (A) sunflower oil; (B) tetradecane; (C) benzyl salicylate; (D) lilial; (E) citronellol; (F) phenoxyethanol. The scaling mark is 20 μm. The thin arrows denote small water drops dispersed in the oil inside the capsule. The thick arrows denote single big water drops located in the oil-filled capsule.

In the case of sunflower oil (Fig. 4A), microcapsules were formed, but their coverage with the silica particles was not so dense. In the case of tetradecane (Fig. 4B), the droplets were much better covered with particles, however some empty spots in the coverage were still seen. As already reported, with limonene dense and robust shells of particles are formed; see e.g. Fig. 2E, F and G. In the case of benzyl salicylate (Fig. 4C), two kinds of structures were simultaneously formed: (i) microcapsules resembling those with limonene, but containing small water droplets dispersed in the oily core and (ii) more rarely – water drops covered with a shell from particles and oil. Structures of the second type were also seen in the case of lilial (Fig. 4D) but in this case the shells from particles + oil were thicker. In the case of citronellol (Fig. 4E), robust capsules with dense particle coverage were formed like those with limonene.

The structures observed in Figs. 4C and 4D can be considered as a manifestation of one-step production of double emulsions; see the detailed review by Clegg et al. [59]. Our micrographs indicate that we are dealing with a hybrid case, in which the outer emulsion is particle-stabilized, whereas the inner emulsion is stabilized by the surfactant (+ polymer).

In some of the experiments with benzyl salicylate, we added blue dye (BD) to the aqueous phase, which led to disappearance of the small water droplets inside the oil microcapsules; the latter were covered very well with particles. In the case of lilial, we observed disappearance of the water drops inside the oil ones (Fig. 4D) when adding another dye – methylene blue (MB). These dyes (BD and MB) are water soluble and do not affect the oil/water interfacial tension. Insofar as each of them contains several aromatic rings, their effect could be attributed to an interaction with the double C=C bond of oleate (the used surfactant), analogous to the π−π stacking.

With other oily phases, viz. silicone oil; light mineral oil; oleic acid; linalool; phenoxyethanol, and benzyl alcohol, it was impossible to produce microcapsules by using the same procedure of homogenization and the same composition of the aqueous phase. In particular, under these conditions, it was impossible to disperse the silicone and mineral oils to drops covered by particles. In contrast, the oleic acid and linalool were dispersed to small emulsion drops, comparable by size with the silica particles (3-4 μm), so that core-in-shell structures could not be formed; see Fig. A6a in the Appendix. The phenoxyethanol formed bigger drops, but they were not covered with silica particles (Fig. 4F). Finally, the benzyl alcohol formed some drop-like structures with aggregated particles around them, but most of the particles were not attached to drops; see Fig. A6b in the Appendix.

The fact that microcapsules are formed with some of the oils and are not formed with other oils can give information about the interactions and mechanism of encapsulation for the investigated system. In Table 1, we have ordered the oils with respect to the increase of their water solubility. As seen in this table, stable capsules are obtained if the oil has intermediate water solubility. Stable capsules were not obtained with oils that have either *zero* or *high* water solubility (with SFO being exception).

We measured also the interfacial tensions of oils of zero, intermediate and high water solubility; see Table 2. The pendant/buoyant drop method with drop-shape analysis was used. Because the basic solution was turbid (the drop was not seen), it was diluted 10 times. For the oils of zero and intermediate water solubility, the addition of surfactant and polymer in the aqueous phase leads to a marked decrease of interfacial tension. In contrast, for the oils of high solubility the difference (between the cases with and without surfactant + polymer) is of the order of the experimental error. This indicates a negligible adsorption of surfactant and polymer at the oil/water interface in the case of oils of high water solubility.

**Table 1.** Solubility of the oils/fragrances in water vs. formation of stable capsules

| Oil | Water solubility | Stable capsules |
|---|---|---|
| Silicone oil, 10 cSt | $\approx 0$ | No |
| Light mineral oil | $\approx 0$ | No |
| Oleic acid | $\approx 0$ | No |
| SFO | $\approx 0$ | Yes |
| Tetradecane | $3.3 \times 10^{-3}$ mg/L | Yes |
| Limonene | 13.8 mg/L | Yes |
| Benzyl salicylate | 24.6 mg/L | Yes |
| Lilial | 33-38 mg/L | Yes |
| Citronellol | 300-307 mg/L | Yes |
| Linalool | 1600 mg/L | No |
| Phenoxyethanol | $\approx 26000$ mg/L | No |
| Benzyl alcohol | $\approx 40000$ mg/L | No |

**Table 2.** Equilibrium interfacial tension (IFT), oil/water vs. oil/solution, where the solution is 0.3 mM KOleate + 0.004 wt% Carbopol-971P + 50 mM NaCl in water.

| Oil | Oil/water IFT (mN/m) | Oil/solution IFT (mN/m) |
|---|---|---|
| Light mineral oil | 52.4 | 40.5 |
| Limonene | 27.8 | 9.8 |
| Benzyl salicylate | 29.0 | 27.0 |
| Linalool | 11.0 | 10.8 |
| Phenoxyethanol | 3.4 | 3.5 |

## 4. Polymer and surfactant interactions at the interface studied by experiments with emulsion films

To get additional information about the differences between oils of different water solubility with respect to their encapsulation, we carried out experiments with oil/water/oil emulsion films by using the SE cell. The evolution of the film was monitored and the film lifetime was registered. Experiments were carried out with representatives of the oils of very low, intermediate and high water solubility (Table 1), viz. light mineral oil, limonene and linalool. Their viscosities are, respectively, 13.2, 0.897 and 4.47 mPa·s at 25 ºC. For each system, at least three independent runs were carried out.

*4.1. Water phase with polymer and salt, without surfactant*

In this series of experiments, the water phase contained 0.04 wt% Carbopol 971P and 500 mM NaCl, without added surfactant. The goal was to detect indications about the polymer attachment to the oil/water interface.

Fig. 5A shows that in the case of *light mineral oil* the film initially contains a large dimple, as indicated by the Newton interference rings. The so called "dimple" (with respect to the oil phase) in fact represents a doubly convex lens filled with the film (aqueous) phase; see, e.g., Ref. [60] for the stages of film evolution. The liquid in this lens gradually flows out leaving behind a plane parallel film of uniform thickness. The latter is, however, unstable and breaks. The total film lifetime is 1 – 2 min. There are no indications for polymer interaction with the film surfaces.

Fig. 5B shows analogous data for the case of *limonene*. Initially, dimple is present in the film. It flows out leaving behind a planar film that contains many dark spots. The latter can be identified as polymer aggregates captured in the film. This may happen only in the case of adsorption of polymer segments at the interface, because otherwise the drainage of water out of the film would drive the polymer in the bulk. In this case, the film lifetime is ≲ 30 s, shorter than for the light mineral oil, which could be explained with faster film drainage because of the lower viscosity of limonene.

Fig. 5C illustrates the results for *linalool*. Dimple is present again. The film lifetime is very short, ≈ 10 s. There is no time for the dimple to flow out. The film breaks at the dimple periphery, where the distance between the two film surfaces is the smallest. The film lifetime is shorter than that for limonene despite the greater viscosity of linalool. This could be explained with dynamic effects engendered by the continuing dissolution of linalool in the aqueous phase (the water solubility of linalool is 116 times higher than that of limonene). Some isolated round-shaped structures inside the dimple (Fig. 5C) could be identified with polymer aggregates. They are more likely closed in the lens-shaped dimple, rather than attached to its surfaces.

The most important results from this series of experiments are that the polymer aggregates attach to the limonene/water interface, whereas they do not attach to the mineral oil/water interface – compare Figs. 5A and B. The pronounced instability of the films with linalool also indicates the lack of polymer adsorption on the linalool/water interface (Fig. 5C).

(A) Oil phase: Light mineral oil; film lifetime = 1 – 2 min

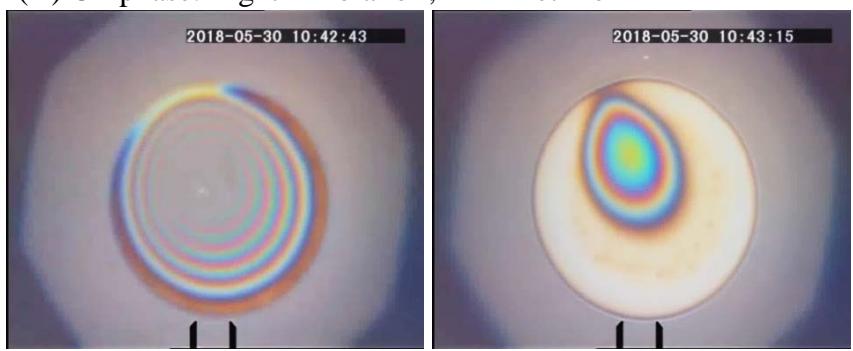

(B) Oil phase: Limonene; film lifetime ≤ 30 s

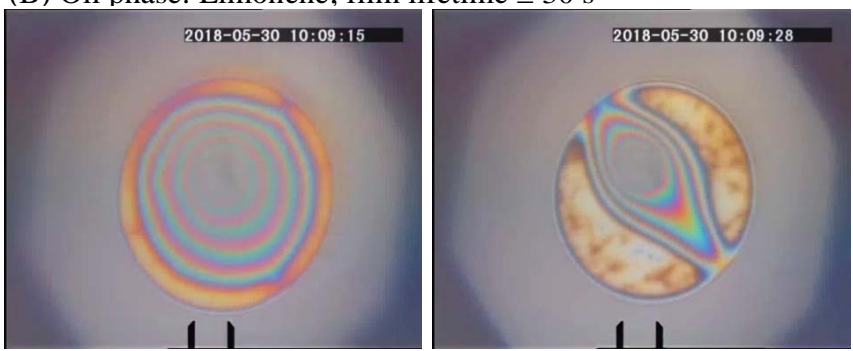

(C) Oil phase: Linalool; film lifetime ≈ 10 s

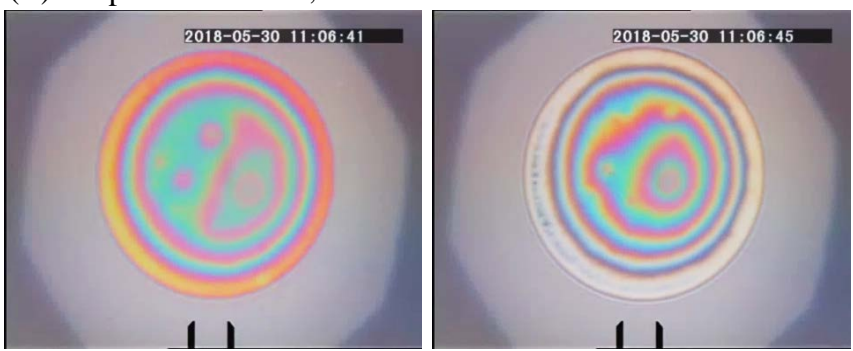

**Fig. 5.** Micrographs of oil/water/oil films in SE cell *without* surfactant in the aqueous phase, which contains 0.04 wt% Carbopol 971P and 500 mM NaCl. The kind of the oil phase and the mean film lifetime are denoted in the figure. For each kind of oil, the first micrograph shows the film just after its formation, whereas the second micrograph – just before its rupture. The scaling mark is 50 μm.

*4.2. Water phase with polymer, salt and surfactant*

In this series of experiments, the water phase contained the full formulation, viz. 0.04 wt% Carbopol 971P, 500 mM NaCl and 3 mM KOleate. The goal of these experiments is to get additional information about the surfactant–polymer interactions at the oil/water interface. The results are illustrated with micrographs in Fig. 6, which show that in the presence of surfactant the film behavior is rather different from that in the absence of surfactant (Fig. 5).

(A) Oil phase: Light mineral oil; stable film (lifetime >20 min)

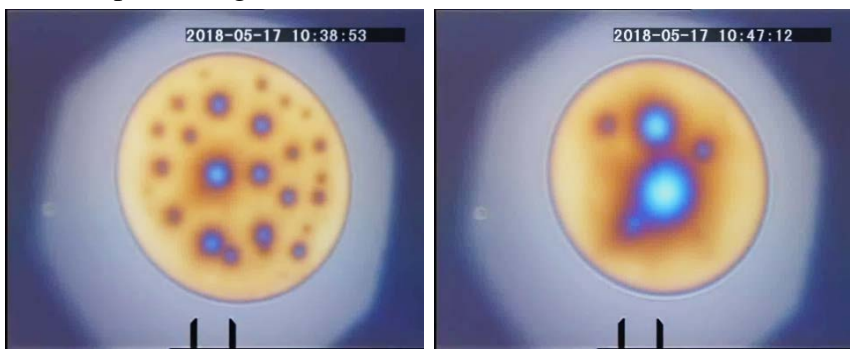

(B) Oil phase: Limonene; lifetime ≈ 2 min

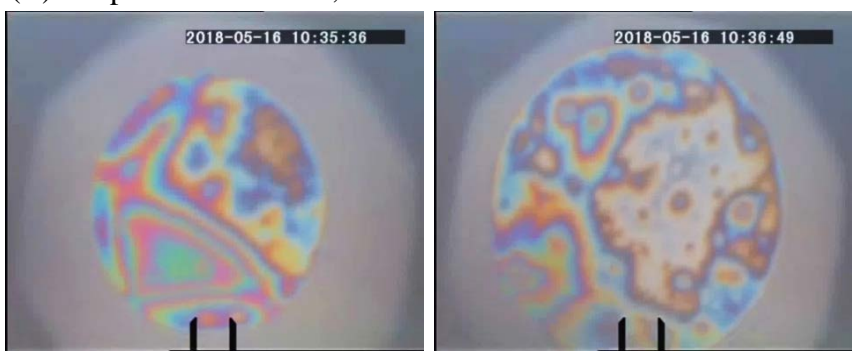

(C) Oil phase: Linalool; lifetime = 10 – 20 s

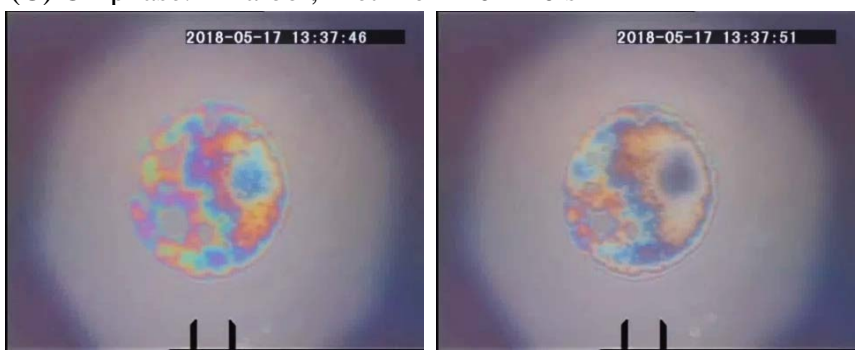

**Fig. 6.** Micrographs of oil/water/oil films in SE cell *with* surfactant in the aqueous phase, which contains 0.04 wt% Carbopol 971P, 500 mM NaCl and 3 mM KOleate. The kind of the oil phase and the mean film lifetime are denoted in the figure. For each kind of oil, the first micrograph shows the film just after its formation, whereas the second micrograph – just before its rupture. The scaling mark is 50 μm.

Fig. 6A illustrates the results in the case with *light mineral oil*. The dimple flows out quickly and a planar film is formed, which contains many entrapped water lenses. Remarkably, this film is stable – it did not break during the whole period of observation, which was 20 min. For all experiments with liquid films reported in the present article, this is the only case with stable film. It is worthwhile noting that in the absence of surfactant (Fig. 5A) and in the absence of polymer (Fig. A7a in the Appendix), the films with light mineral oil are unstable. The stability of the films in Fig. 6A indicates the presence of surfactant-polymer interaction. The most probable structure implied by the experimental observations is that the surfactant adsorbs at the oil/water interface and the polymer adsorbs on the surfactant headgroups, thus forming robust adsorption layers at the film surfaces, which stabilize the emulsion film.

Fig. 6B illustrates the results in the case with *limonene*. The irregular interference pattern with separate round spots indicates film of uneven thickness, which contains trapped polymer aggregates. This experimental pattern could be due to competitive adsorption of surfactant and polymer at the oil/water interface. The observed pattern is dynamic, which could be explained with simultaneous film drainage and limonene dissolution in the aqueous phase. Eventually, the film breaks; its average lifetime is ≈ 2 min, which is ca. 4 times longer than in the case without surfactant (KOleate).

Fig. 6C illustrates the case with *linalool*. In this case the dynamics is the fastest – the dimple quickly flows out; the formed film has uneven thickness that fluctuates with time and the film breaks within 10 – 20 s. The obvious difference between the patterns in Fig. 5C and 6C could be explained with the fact that in the former case (no surfactant) the faster dissolution of linalool from the film surfaces in the film interior gives rise to osmotic pressure that counterbalances the Laplace pressure of the convex meniscus and, thus, stabilizes the dimple. With surfactant, this effect is suppressed; the dimple quickly flows out but the formed film is rather unstable and ruptures.

The most important results from this series of experiments are the indications (i) for surfactant-polymer interaction (Fig. 6A); (ii) that the surfactant cannot displace the polymer from the limonene/water interface (Fig. 6B), and (iii) that the surfactant adsorbs at the linalool/water interface and slows down the dissolution of linalool in the water.

## 5. Discussion

Here, our goal is to identify the main factors that favor the formation of stable microcapsules by analyzing the experimental results. As defined in Section 2.2, we call 'stable' the capsules, which are not destroyed after rinsing with water.

In the absence of surfactant and polymer, oil drops covered with particles can be formed (Fig. 2A and B), but they are destroyed by weak shaking of the vial with the suspension. This could be explained with the fact that the hydrophilic particles (with small contact angle measured across water; Fig. 7A) are weakly attached to the drop surfaces, so that the hydrodynamic flow detaches them and destabilizes the emulsion.

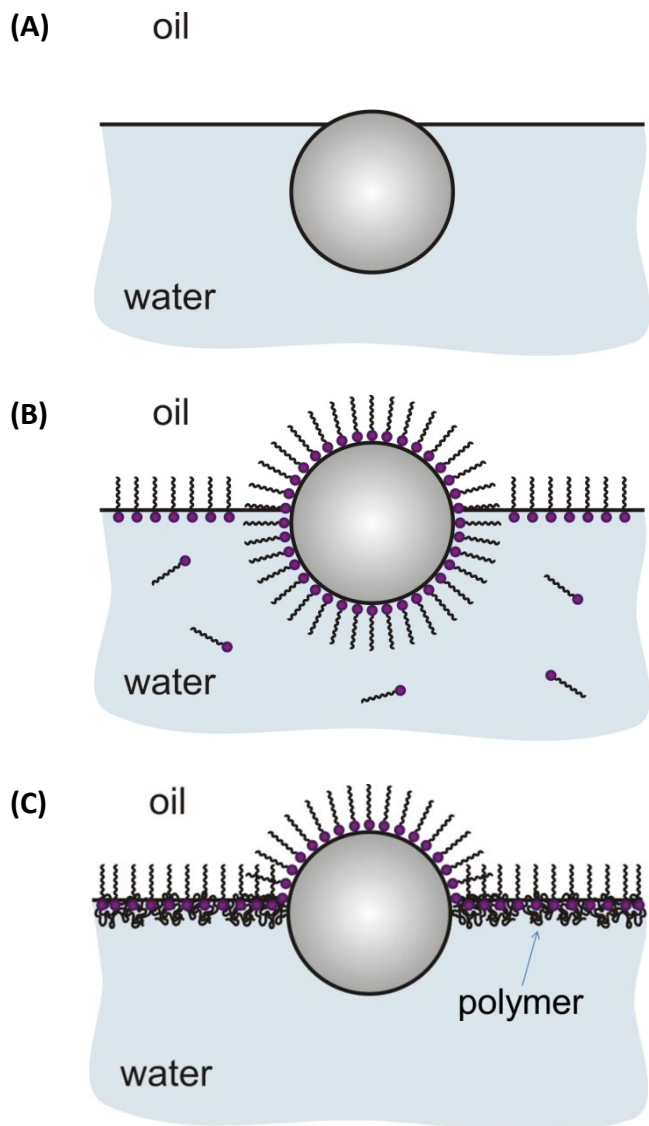

**Fig. 7.** (A) The used hydrophilic particles have small contact angle across water and can be detached from the oil/water interface by hydrodynamic flows. (B) The surfactant (potassium oleate) hydrophobizes the particles and stabilizes their attachment to the liquid interface. (C) The irreversible adsorption of polymer at the oil/water interface blocks (i) the dissolution of the oil/fragrance in the water, and (ii) surfactant desorption into the aqueous phase when replaced with pure water after rinsing. (The relative size of the surfactant molecules is much smaller than drawn in the figure.)

If surfactant is added, the formed emulsion drops are covered by adsorbed particles (Fig. 2C and D) and the respective emulsion is *stable* upon shaking. However, after replacing the surfactant solution with pure water, the emulsion is destabilized. This can be explained with fact that the surfactant hydrophobizes the silica particles and leads to their deeper entry at the oil/water interface (Fig. 7B), i.e. to stronger particle attachment to the drop surfaces. However, the rinsing with water removes the adsorbed surfactant; the particles recover their hydrophilicity and acquire the configuration in Fig. 7A, so that the microcapsules become unstable upon shaking.

In the presence of surfactant and polymer (Fig. 2E, F, G and H), the formed capsules are stable after rinsing with water for a period of at least 8 months. This could be explained with the irreversible adsorption of polymer at the oil/water interface. The layer of irreversibly adsorbed polymer blocks desorption of surfactant from the oil/water and particle/oil interfaces. (Otherwise, the particles would become hydrophilic and desorb in the water phase, which is not the case; see Fig. A8 in the Appendix.) Thus, the particles remain strongly attached to the drop surfaces and the spaces between them remain covered with a mixed polymer + surfactant layer (Fig. 7C). The latter mixed layer suppresses (decelerates) also the dissolution of oil in the water phase, where the oil (e.g. limonene) exhibits a limited solubility. This is supported by the fact that in aqueous environment the microcapsules remain intact at least for 8 months.

The microcapsule stabilization model illustrated in Fig. 7C, involves energetically favorable interactions (attractive forces) between (i) particles and surfactant; (ii) surfactant and oil; (iii) polymer and oil, and (iv) polymer and surfactant; see Fig. 8.

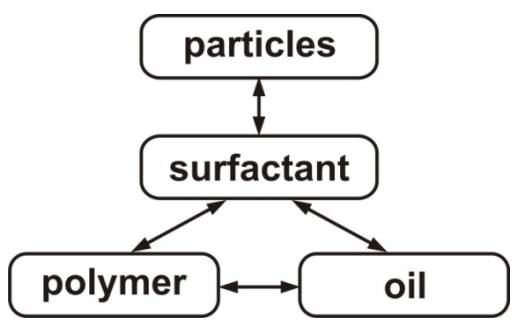

**Fig. 8.** Sketch of the interactions between the system's ingredients, which should exist in order stable microcapsules to be formed; the arrows symbolize energetically favorable (attractive) interactions. Because the silica particles are hydrophobized by an adsorption layer of surfactant molecules, the particle-oil and particle-polymer interactions are mediated by the surfactant, as schematically shown in the figure.

Indeed, the *particle–surfactant* interaction is crucial for the particle hydrophobization and firm attachment to the oil/water interface (Fig. 7B). As already mentioned, the surfactant (KOleate) concentration should not be too high – otherwise, the particles become too hydrophobic and begin to aggregate and even to enter the oily phase; see Fig. A2 in the Appendix.

The *surfactant–oil* interaction is always important for the emulsification: the surfactant molecules adsorb much faster to the oil drops (than the colloidal particles) and prevent the coalescence of the formed drops upon the collisions between them during emulsification.

The *polymer-oil* interaction is crucial for the irreversible adsorption of the polymer at the oil/water interface. Otherwise, the polymer would be washed out upon rinsing and the capsules would be destroyed. The reason for the capsule destruction at alkaline pH is that the polymer (Carbopol or PAA) becomes ionized and more hydrophilic. For this reason, the rinsing with water of alkaline pH leads to polymer desorption from the oil/water interface, followed by surfactant desorption, particle hydrophilization and detachment, and coalescence of the unprotected emulsion drops.

Finally, the *polymer-surfactant* interactions allow retention of surfactant molecules at the oil/water interface, and more importantly – at the particle/oil interface, which ensures sufficiently large contact angle and firm particle attachment to the surfaces of the emulsion drops. This makes possible the formation of the mixed polymer-surfactant layer depicted in Fig. 7C.

The destruction of the capsules by other surfactants like Tween 20, C12-SME and SLES-1EO at sufficiently high concentrations (see Section 3.4) can be explained with the energetically-advantageous formation of complexes of these surfactants with the polymer in the bulk, which leads to destruction of the protective polymer-surfactant layer at the surface of the microcapsules and to their destabilization.

In the case of oils of *very low* water solubility (like the silicone and mineral oils) microcapsules, which are stable upon rinsing with water, were not formed (Table 1). This fact indicates that for the formation of stable capsules it is necessary not only the surfactant, but also (segments of) the polymer to adsorb at the oil/water interface. Thus, the polymer can anchor the mixed (polymer + surfactant) adsorption layer to the oil water interface and to prevent its desorption upon rinsing. However, in the case of water-insoluble oils the anchoring effect of polymer is missing (see Fig. 5A and the related text), so that the microcapsules are destabilized upon rinsing because the surfactant desorbs from the oil/water interface, together with the polymer. Indeed, as deduced from Fig. 6A, the surfactant mediates the polymer adsorption to the interface of insoluble oils and, consequently, the surfactant desorption triggers also polymer desorption.

The case of SFO is more special, because this oil typically contains an admixture of ca. 1% fatty acids [61], which can favorably interact with the polymer and ensure its adsorption at the oil-water interface. This could explain the formation of stable capsules with SFO despite its extremely low water solubility (Table 1).

In the case of oils that exhibit considerable water solubility, like phenoxyethanol, the adsorption of polymer + surfactant at the oil/water interface is too weak (as indicated by the interfacial tension – see Table 2). Hence, in this case the encapsulation mechanism illustrated in Fig. 7C cannot be realized.

The oleic acid is another special case – under the same emulsification procedure it forms too small droplets, which are comparable by size with the used silica particles, so that formation of core-in-shell structures is impossible; see additional information in the Appendix.

## 6. Conclusions

Colloidosomes provide a possibility to encapsulate oily substances in water in the form of core-in-shell structures. However, if the oil exhibits a finite solubility in water, it is released in the aqueous phase through the openings between the particles of the colloidal shell. In this study, we produced microcapsules with shell from colloidal particles, where the interparticle openings are blocked by mixed layers from polymer and surfactant (Fig. 7C). In other words, if the particles are analogs of bricks, the layers from polymer + surfactant play the role of mortar.

For this goal, we used hydrophilic silica particles of diameter 3 – 4 μm, which were partially hydrophobized by the adsorption of potassium oleate, so that they were able to stabilize Pickering emulsions. The procedure of encapsulation is simple and includes single homogenization by ultrasound. Various polymers were tested (Section 3.3). The most stable microcapsules were obtained with Carbopol® 971P. They have a mean diameter of about 20 μm and remained stable after 8 months shelf storage at room temperature. The produced microcapsules are stable when rinsed with water of pH in the range 3 – 6. However, if they are dispersed in water of pH > 6, they are destabilized and release their cargo, i.e., they are pH responsive.

With the optimized formulation of silica particles, polymer, potassium oleate and NaCl, we were able to encapsulate various oils and fragrances, such as tetradecane, limonene, benzyl salicylate and citronellol. All of them have a limited solubility in water. In contrast, no stable microcapsules were obtained with oils that either have zero water solubility (mineral and silicone oil) or high water solubility (phenoxyethanol and benzyl alcohol); see Table 1. By analysis of results from additional interfacial-tension measurements and thin-film experiments, we concluded that a key factor for obtaining stable capsules is the irreversible adsorption of the polymer at the oil/water interface in the openings between the particles.

The role of the adsorbed polymer is not only to block the dissolution of the encapsulated oil/fragrance in the aqueous phase, but also to prevent dissolution of the surfactant molecules, which hydrophobize the particles and stabilize them at the oil/water interface. Upon rise of pH above 6, the increased negative electric charge of the polymer leads to its desorption into water. This frees the way for the hydrophobizing surfactant molecules on the particle/oil interface (Fig. 7C) to migrate toward water. As a result, the particles recover their hydrophilicity and detach from the oil/water interface, so that the capsule is destabilized. In other words, not only the polymer, but also the surfactant, plays an important role for the pH-responsiveness of the produced microcapsules.

The obtained information about the role of various factors for the stabilization/destabilization of microcapsules, which are based on the brick-and-mortar concept, can be further used to achieve better capsule stabilization; selection of polymers that are appropriate for different classes of oils, as well as for the production of smaller capsules stabilized by nanoparticles.

**Acknowledgments**

The authors gratefully acknowledge the support from the National Science Fund of Bulgaria, Grant No. DN 09/8/2016. They thank Dr. Simeon Stoyanov for the helpful discussions; Dr. Rumyana Stanimirova and Ms. Veronika Ivanova, MSc, for the interfacial tension measurements, and Ms. Varbina Ivanova for the ζ-potential measurements.**Appendix. Supplementary material**

Supplementary data associated with this article can be found in the online version, at http://dx.doi.org/10.1016/j.colsurfa.2018....

# Appendix. Supplementary material

for the article

## Encapsulation of oils and fragrances by core-in-shell structures from silica particles, polymers and surfactants: The brick-and-mortar concept

Gergana M. Radulova [a], Tatiana G. Slavova [a], Peter A. Kralchevsky [a,*], Elka S. Basheva [a], Krastanka G. Marinova [a], Krassimir D. Danov [a]

[a] *Department of Chemical and Pharmaceutical Engineering, Faculty of Chemistry and Pharmacy, Sofia University, 1164 Sofia, Bulgaria*

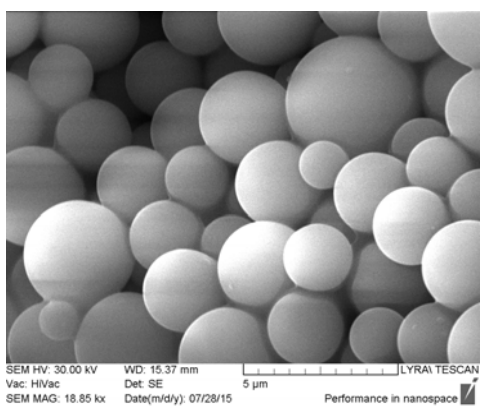

**Fig. A1.** Micrograph of the used ECELICA $SiO_2$ particles taken by scanning electron microscope.

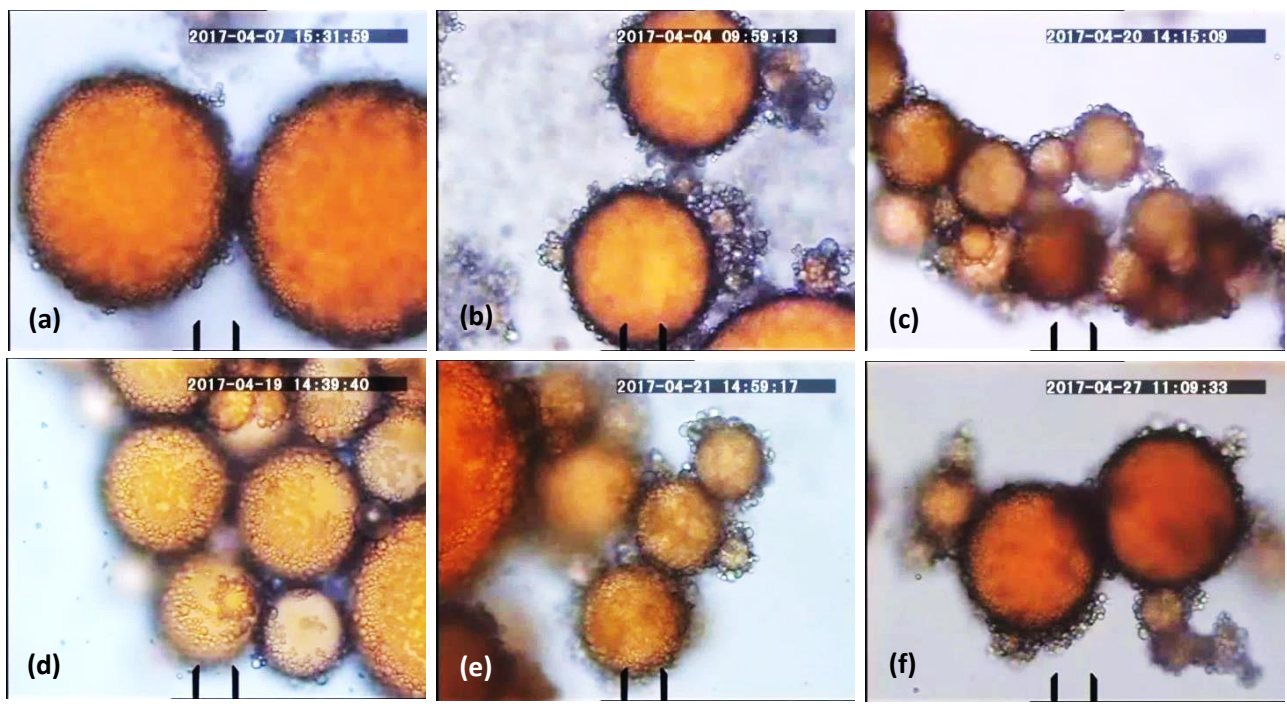

**Fig. A2**. Effect of electrolyte (no surfactant; no polymer). Micrographs of particle stabilized 10 vol % limonene-in-water emulsions, for which the aqueous phase contains 5 wt% silica particles and various dissolved electrolytes: (a) KCl; (b) NaCl; (c) $ZnCl_2$; (d) $CaCl_2$; (e) $FeSO_4$, and (f) $AlCl_3$. The scaling mark is 20 μm.

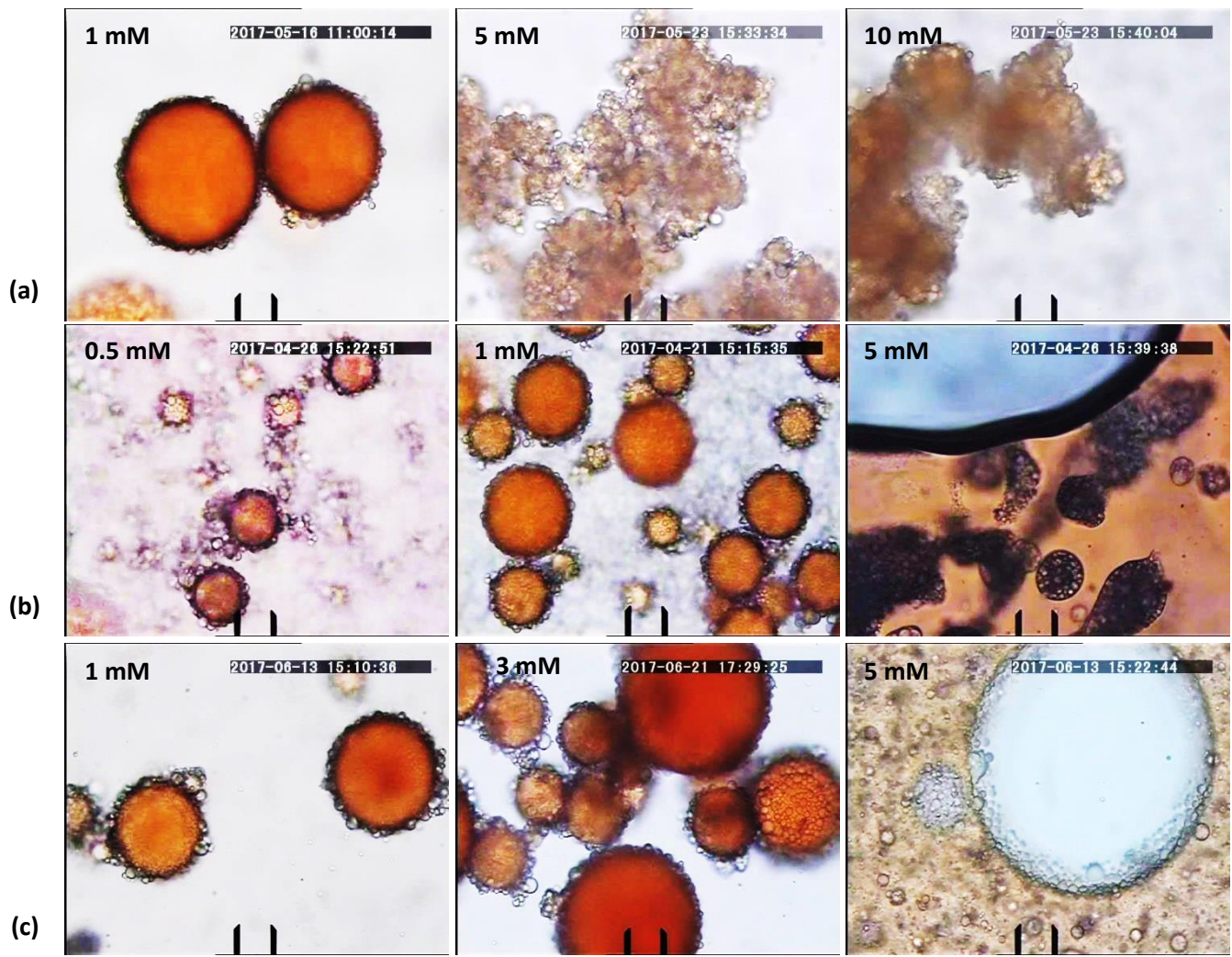

**Fig. A3**. Effect of surfactant (no polymer). Micrographs of the dispersions from 10 vol% limonene and aqueous phase containing 5 wt% silica particles; 500 mM NaCl, and 3 mM carboxylate, which is: (a) sodium laurate; (b) potassium myristate, and (c) potassium oleate. The scaling mark is 20 μm.

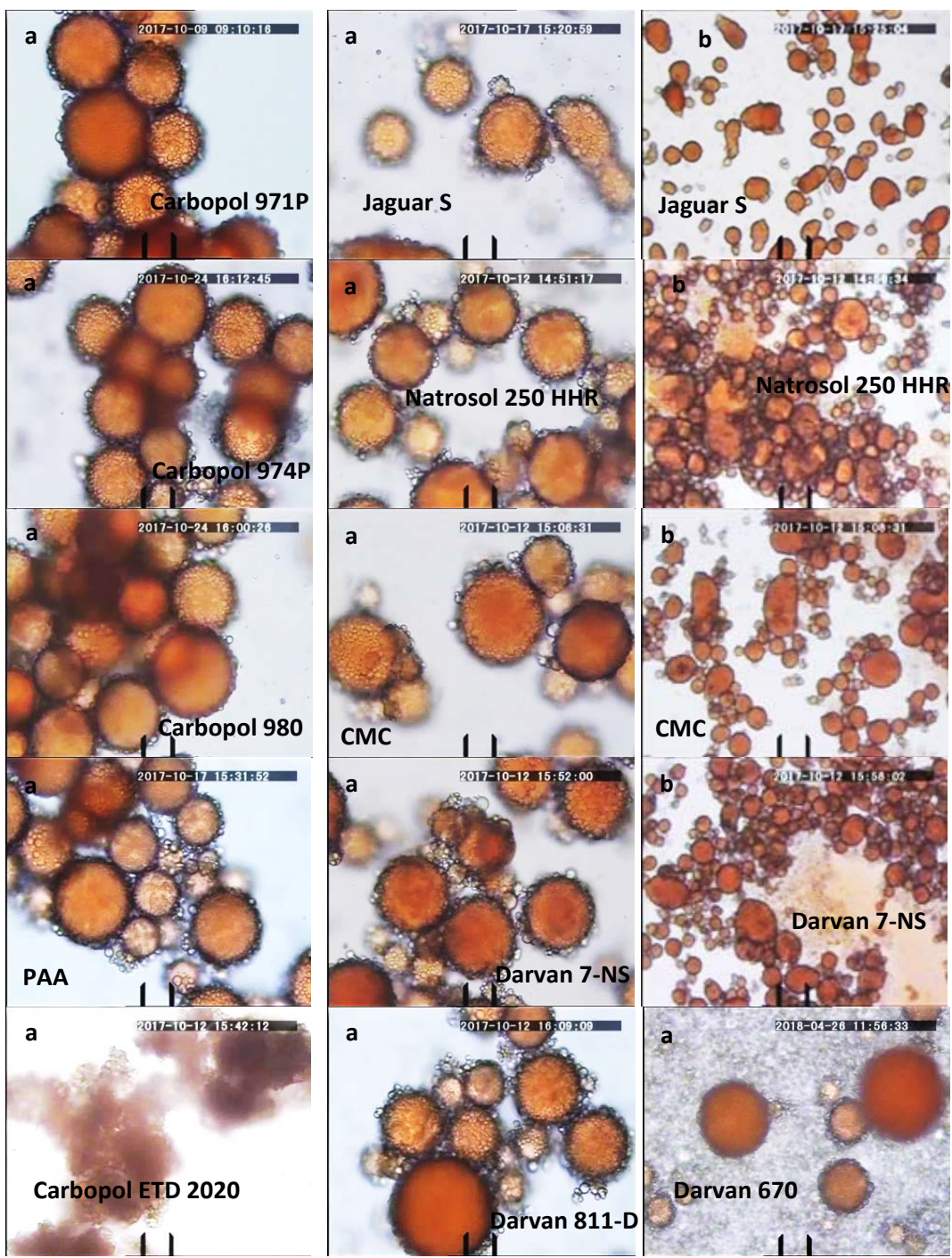

**Fig. A4**. Effect of the kind of polymer. Micrographs of the dispersions from 10 vol% limonene and aqueous phase containing 5 wt% silica particles; 500 mM NaCl; 3 mM KOleate, and 0.04 wt% polymer of kind that is given in the figure. The scaling mark is 20 and 50 μm for the micrographs denoted, respectively, with "a" and "b".

| pH = 3 – 3.5 | pH = 5.5 – 6 | pH = 10 – 10.5 |

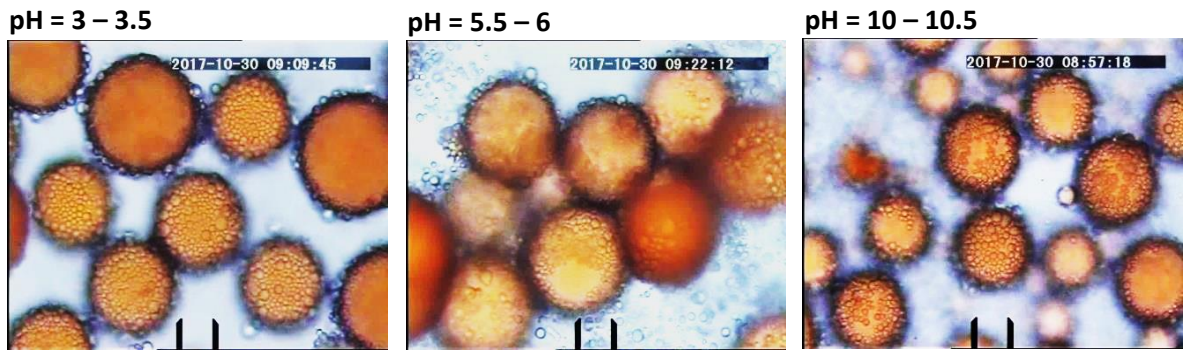

**Fig. A5.** Effect of temperature. Samples at different pH values (denoted in the figure) were heated at 80 °C for 1 hour. The composition of the dispersion is 10 vol% limonene; 5 wt% silica particles; 500 mM NaCl; 3 mM KOleate, and 0.04 wt% Carbopol 971P. The scaling mark is 20 μm.

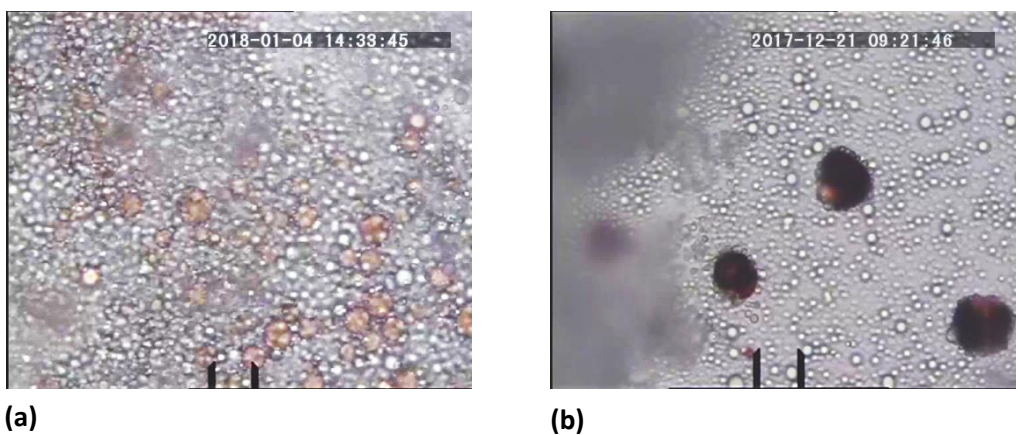

(a)            (b)

**Fig. A6.** Effect of the kind of oil. Micrographs of dispersion with the same composition as in Fig. A4, but instead of limonene, the oil phase is (a) linalool and (b) benzyl alcohol. The scaling mark is 20 μm.

**Film phase: water solution of 3 mM KOleate + 500 mM NaCl (no polymer)**

(a) Foam film: The outer phase is air; film lifetime: ≈15 s

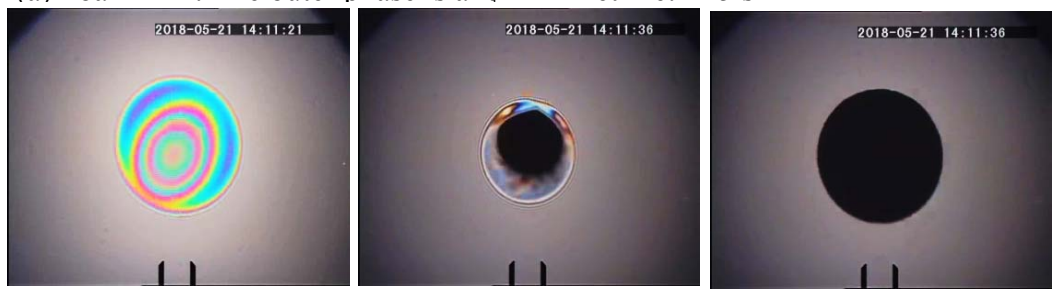

(b) Oil phase: Light mineral oil; film lifetime ≈ 1 min

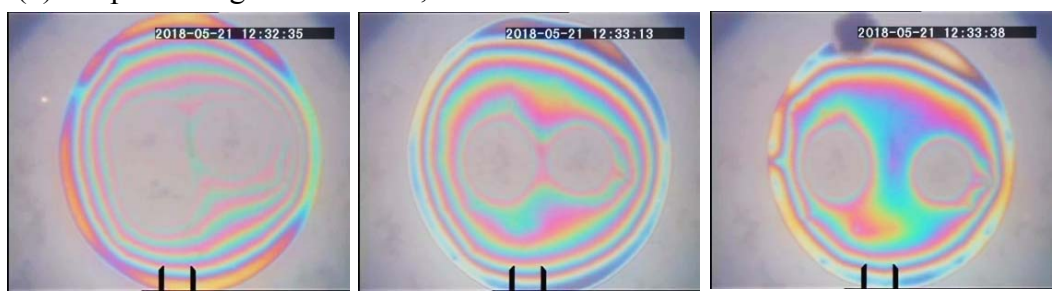

(c) Oil phase: Limonene; lifetime ≤ 1 min

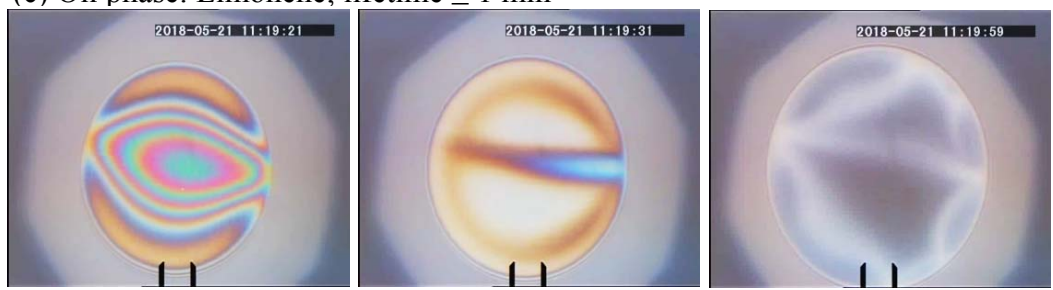

(d) Oil phase: Linalool; film lifetime ≤ 2 min

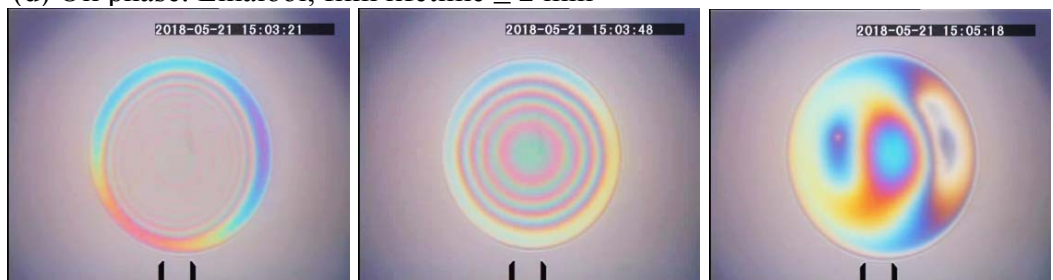

**Fig. A7.** Micrographs of thin aqueous films in SE cell *without* polymer in the aqueous phase, which contains 3 mM KOleate and 500 mM NaCl. The leftmost micrograph shows the film just after its formation, whereas the rightmost micrograph – just before its rupture; the micrograph in the middle illustrates an intermediate stage of film evolution. (a) In the case of foam film, a thin black film is formed, which is unstable and quickly ruptures. (b) In the case of light mineral oil, the film forms a dimple and thins slower because of the higher viscosity of the oil phase (13.2 mPa·s); the film breaks soon after the appearance of a darker spot at its periphery. (c) In the case of limonene, which has a lower viscosity (0.897 mPa·s), the film thins faster and quickly ruptures. (d) The linalool has an intermediate viscosity (4.47 mPa·s); the film lives slightly longer because of the stabilizing effect of the mass transfer accompanying the linalool dissolution from the film surfaces into water. The scaling mark is 50 μm.

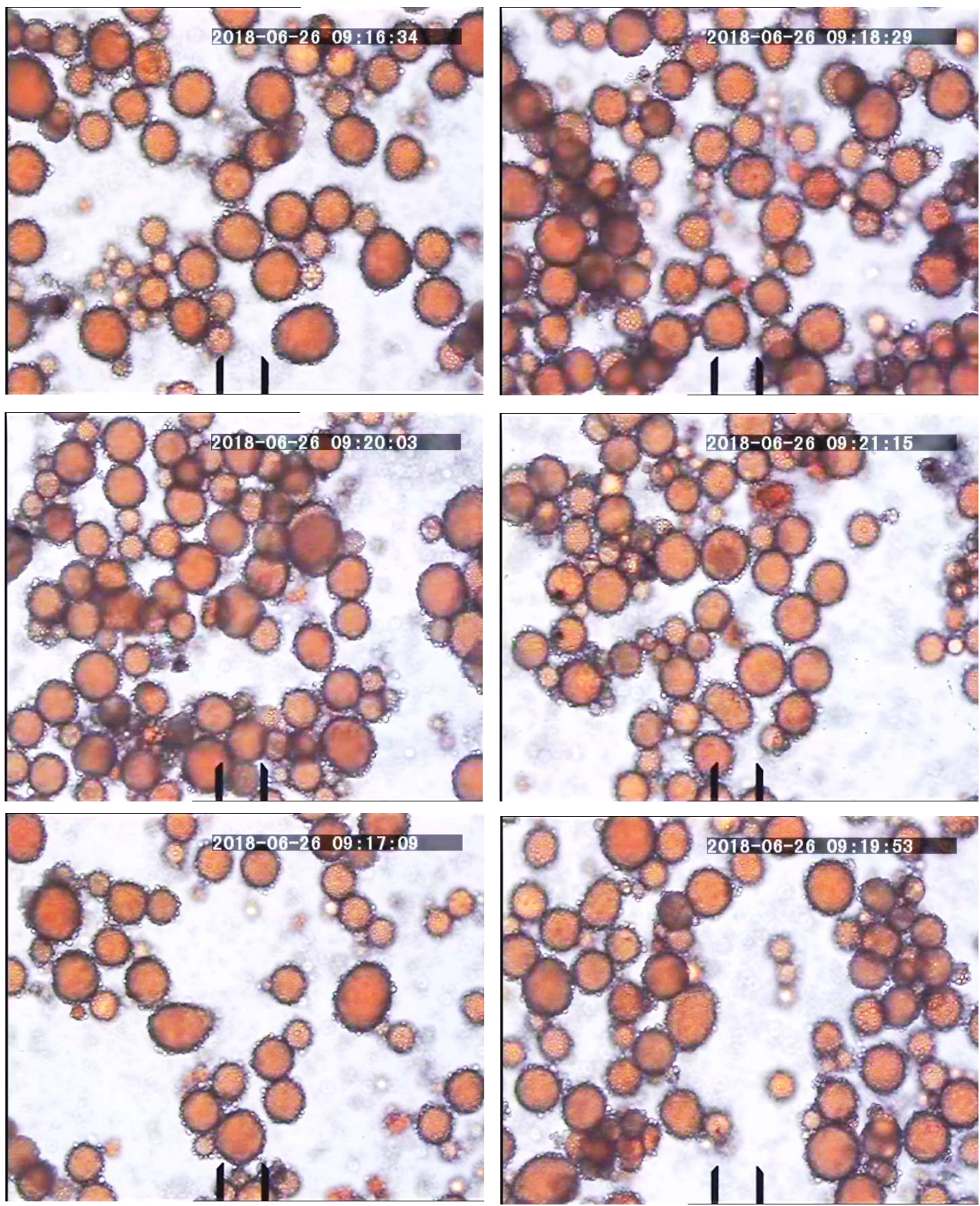

**Fig. A8.** Stability of the produced microcapsules upon shelf storage for 8 months in water of pH = 5.5 at 25 ºC. The capsules were prepared from a dispersion containing 10 vol% limonene; 5 wt% silica particles; 3 mM KOleate; 0.04 wt% Carbopol 971P, and 500 mM NaCl. Preparation date: 20/10/2017. The micrograph, taken on 26/06/2018, indicate that the capsules are stable after storage for more than 8 months. The scaling mark is 50 μm.

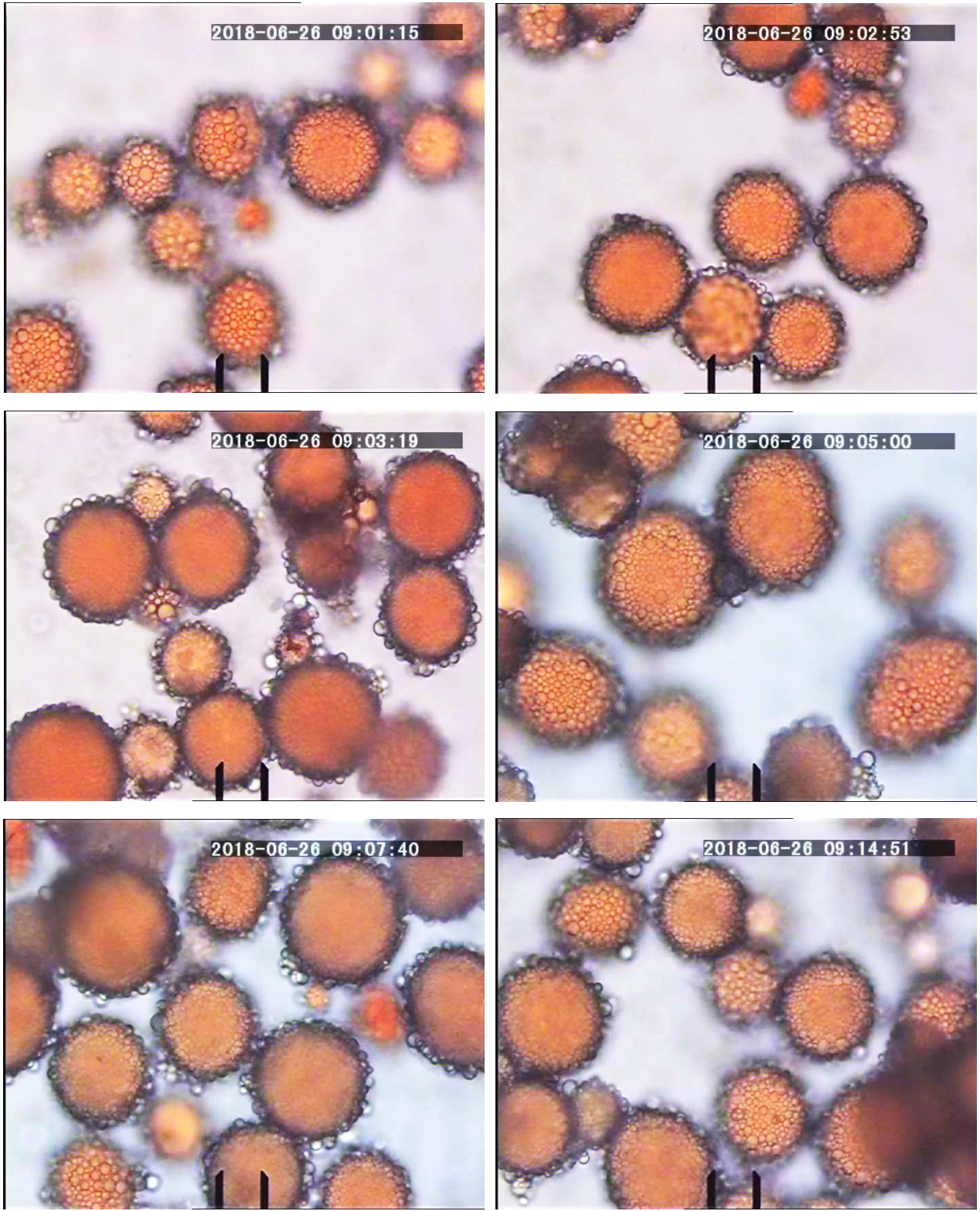

**Fig. A9.** The same as in Fig. A8, but at a higher magnification – the scaling mark is 20 μm.

**Table A1.** Viscosity ranges of Carbopol solutions, as provided by the producer and zeta potentials of Carbopol aggregates in water measured by us.

| Polymer | Viscosity, mPa·s (0.5 wt% polymer at pH 7.5) | $\zeta$-potential, mV (0.1 wt% polymer + 10 mM NaCl at pH 7.5) |
|---|---|---|
| Carbopol 971P | 4,000 – 11,000 | −73 |
| Carbopol 974P | 29,400 – 39,400 | −47 |
| Carbopol 980 | 40,000 – 60,000 | −58 |
| Carbopol ETD 2020 | 47,000 – 77,000 | −51 |

In Table A1, the viscosity ranges of Carbopol solutions (as provided by the producer) give indications for $M_w$, whereas the zeta potentials of Carbopol aggregates in water (measured by us), give indications for the degree of crosslinking. The higher is the degree of crosslinking the lower is the magnitude of the negative zeta potential. (This is because two ionizable carboxylic groups disappear upon a crosslink formation.) Hence, from the data in Table A1, it follows that Carbopol 971P is the polymer of the lowest molecular mass and the lowest degree of crosslinking among the used Carbopols.

The *oleic acid* is a special case – under the same emulsification procedure it forms too small droplets, which are comparable by size with the used silica particles, so that formation of core-in-shell structures is impossible. The reason for the formation of so small emulsion drops could be that the emulsification is accompanied with a chemical reaction – saponification of oleic acid molecules exposed to contact with water at the drop surfaces. Indeed, the dissolved KOleate raises the pH to 8 – 9, at which the oleic acid molecules undergo hydrolysis. Under these conditions, one could form core-in-shell structures with smaller particles (nanoparticles), which could be a task for a subsequent study.